\documentclass[aps,preprint,showpacs]{revtex4}
\usepackage{amsmath}
\newcommand{\be}{\begin{equation}}
\newcommand{\ee}{\end{equation}}
\newcommand{\bea}{\begin{eqnarray}}
\newcommand{\eea}{\end{eqnarray}} 
\newcommand{\ba}{\begin{array}}
\newcommand{\ea}{\end{array}}

\newcommand{\bb}{\bibitem}

\begin{document}

\title{\bf Critical exponents from parallel plate geometries subject to 
periodic and antiperiodic boundary conditions }
\author{Jos\'e B. da Silva Jr.\footnote{e-mail:jborba@df.ufpe.br} and 
Marcelo M. Leite\footnote{e-mail:mleite@df.ufpe.br}}
\affiliation{{\it Laborat\'orio de F\'\i sica Te\'orica e Computacional, 
Departamento de F\'\i sica,\\ Universidade Federal de Pernambuco,\\
50670-901, Recife, PE, Brazil}}

\vspace{0.2cm}
\begin{abstract}
{\it We introduce a renormalized $1PI$ vertex part scalar field theory 
setting in momentum space to computing the critical 
exponents $\nu$ and $\eta$, at least at two-loop order, for a layered 
parallel plate geometry separated by a distance L, with periodic as well as 
antiperiodic boundary conditions on the plates. We 
utilize massive and
massless fields in order to extract the exponents in independent ultraviolet 
and infrared scaling analysis, respectively, which are required in a complete 
description of the scaling regions for finite size systems. We prove that 
fixed points and other critical amounts either in the ultraviolet or in the 
infrared regime dependent on the plates boundary condition are a general 
feature of normalization conditions. We introduce a new description of 
typical crossover regimes occurring in finite size systems. Avoiding these 
crossovers, the three regions of finite size scaling 
present for each of these boundary conditions are shown to be 
indistinguishable in the results of the exponents in periodic and antiperiodic 
conditions, which coincide with those from the (bulk) infinite system.}      
\end{abstract}
\vspace{1cm}

\maketitle

\newpage
\section{Introduction}

\par Finite-size effects manifest themselves generically whenever particles 
or fields are confined within a given volume whose limiting surfaces are 
separated by a certain distance $L$. Their size and shape can affect key 
properties of the system in comparison with those obtained from the 
$L \rightarrow \infty$ limit (``bulk system''). Perhaps the most investigated 
aspects are related to critical properties of finite systems \cite{F,FB}, 
where field-theoretic methods can be employed in the vicinity of the phase 
transitions taking place in the system under consideration. Experimentally, 
the simplest realization of such critical behavior and the role played by 
the finite size corrections show up in parallel plate geometries, for 
instance, in coexistence curves of critical films of certain 
fluids \cite{SMMO} as well as superfluid transition features 
(e.g., specific heat amplitudes) in confined 
$ ^{4}$He \cite{Getal1,Getal2}. From the theoretical 
viewpoint, field theory studies have been put forth to explain these 
effects not only for $ ^{4}$He \cite{HD}, but also in thin slabs 
\cite{KD1,KD2} formed by wetting phenomena \cite{Tabo}. The Casimir effect 
has also been investigated in superfluid wetting films \cite{MGS}. Plus, 
the recent study of some microscopic properties of finite-lenght cobalt 
nanowires \cite{GL} reveals that the influence of the finiteness is a 
ubiquitous theme in several properties of physical systems.
\par Momentum space $\epsilon$-expansion description \cite{WF} of 
critical properties for finite size systems was presented some time ago by 
Nemirovsky and Freed ($NF$) \cite{NF1,NF2}. The simplest approach uses a 
parallel plate layered geometry, namely, a (slab) volume of material whose 
limiting surfaces (plates) are of infinite extent along $(d-1)$ 
spatial directions and are separated by a distance $L$. This 
parallelepiped-shaped (e. g., magnetic) material possess a field-theoretical 
description of its critical behavior in momentum space which 
requires continuous 
momenta components parallel to the $(d-1)$ spatial directions and discrete 
``quasimomenta'' along the finite size direction of the material. It is 
basically a combination of effects coming from volume (bulk), finite size and 
surface phenomena. The first two are dominant whenever the absolute value 
of the order parameter (field) is chosen to have the same 
(not specified {\it a priori}) value at the limiting plates. 
(If an external field is allowed in addition to the bulk order 
parameter, and kept at a fixed value at the 
limiting surfaces, surface effects will become proeminent in the discussion 
of the subsequent criticality, beyond the simpler volume (bulk) plus 
finite size corrections pattern.) 
\par These geometric restrictions can be realized as many different boundary 
conditions implemented in the bare free propagator. The above mentioned 
simpler finite size correction shall interest us throughout and can be 
modelled when periodic and antiperiodic boundary conditions are employed. 
(Dirichlet and Neumann boundary conditions mimic free surfaces, are 
appropriate to explain finite size plus surface effects and shall not concern 
us in what follows.) The limitation caused by the boundary conditions provides 
a scaling variable $\frac{L}{\xi_{\infty}}$, where $\xi_{\infty}$ is the 
(bulk) correlation length of the infinite system. Many computations up to 
first order in $\epsilon$ of amplitudes connected to Green functions have 
been carried out within this massive framework, as well as some universal 
amplitude ratios of certain thermodynamical potentials \cite{Leite}. Within 
this context, three scaling regions induced by the limitation have been 
proposed. The first one is characterized by $\frac{L}{\xi_{\infty}}>1$ 
where perturbative methods can be applied and the physics is quasi 
$d$-dimensional, characterized by bulk critical exponents {\it but} 
limitation dependent amplitudes. The second region corresponds to 
$\frac{L}{\xi_{\infty}}\sim 1$ and it was conjectured that the critical 
behavior is neither $d$-dimensional nor $(d-1)$-dimensional. The third region 
is associated to values of the variable $\frac{L}{\xi_{\infty}}<1$. It was 
also argued that in this regime the physics is almost $(d-1)$-dimensional and 
usual perturbation expansions break down \cite{NF2}. Another prediction 
stated that the normalization functions and the exponents would be the same 
as those found in the infinite system for the boundary conditions above 
mentioned. 
\par In this work we introduce a one-particle irreducible ($1PI$) renormalized 
field-theoretic version of the $NF$ formalism in order to investigate finite 
size corrections to normalization functions, fixed points, etc., at higher 
order in a perturbative loop expansion which are 
dependent upon the boundary condition on the plates. Concrete applications for 
periodic ($PBC$) and antiperiodic ($ABC$) boundary conditions are explored 
through the computations of the critical exponents $\eta$ and $\nu$ in finite 
size scaling using the diagrammatic method in momentum space, at least up to 
two-loop order. We improve the understanding of the three scaling regions and 
show that the finite size effects related to the limitation caused by 
the boundary conditions do not show up in the exponents themselves, although 
they modify the ingredients required to compute them. 
\par We utilize massive fields obeying 
these boundary conditions on the plates for nonvanishing values of $L$ 
corresponding to fixed finite values of the bulk correlation length. 
Both first region ($L > \xi_{\infty}$) and the second one associated to finite 
values of $L$ ($\rightarrow \xi_{\infty}$) can be described satisfactorily 
within this massive framework. The remaining region 
is treated with massless fields having infinite bulk correlation length. 
In that case, second region is realized through the limit 
$L \sim \xi_{\infty} \rightarrow \infty$. The third region naturally describes 
arbitrary finite values of $L$ and can only be approached using massless 
fields. The universal results obtained are shown to be valid for the three 
regions determined by the boundedness variable $\frac{L}{\xi_{\infty}}$ which 
interpolates from infinite to {\it finite (not so small) values of $L$}. 
The failure of the finite size phenomenological scaling arguments regarding 
the second and especially the third scaling region is demonstrated for the 
first time.
\par From our analytical expressions described essentially in terms of 
elementary primitives, we demonstrate that the dominant 
contribution of the finite size correction goes with the inverse power of 
$L$ only for periodic boundary condition, where dimensional crossover starts 
to set in the critical behavior. This pattern occurs for both massive ($t>0$) 
{\it and} massless ($t=0$) regimes, although with a larger 
coefficient in the last situation. Antiperiodic boundary conditions have the 
usual crossover at $t<0$ in the massive theory as previously discussed by 
Nemirovsky and Freed \cite{NF2}. Furthermore, our analytical method shows 
clearly the existence of {\it a new type of crossover which takes place for 
ABC at $t>0$ when a term proportional to $lnL$ becomes important for small 
values of $L$}. On the other hand, $ABC$ in the massless regime $t=0$ presents 
a power law of the type $L^{-2}$ whenever $L$ is small. Therefore massless 
and massive crossover regimes are completely different for $ABC$, which is 
demonstrated here for the first time. We show that, as long as we avoid these 
crossover regions for very small values of $L$, there is no breakdown of the 
$\epsilon$-expansion into third region and demonstrate the validity of the 
computation of the exponents. As far as critical exponents are concerned, 
{\it the physics of the systems in the three 
regions is actually quasi $d$-dimensional}, for the bulk critical exponents 
are recovered from the finite size evaluation irrespective of the boundary 
condition and the value of $L$, i.e., independent of the limitation variable 
$\frac{L}{\xi_{\infty}}$. 
\par The paper is organized as follows. In Section II we describe the 
formalism of massive fields with fixed 
finite correlation length (mass). The case 
$L \rightarrow \infty$  corresponding to region a) is shown to 
smoothly reproduce the bulk exponents.  An introduction to the 
$L \rightarrow 0$ limit and  how it is related to dimensional 
crossover is presented as well as the result of the solution to 
higher loop diagrams away from the dimensional crossover region. Section III 
presents the computation of the critical exponents $\eta$ 
and $\nu$ using normalization conditions utilizing the Feynman diagrams 
outlined in the previous section, at least up to two-loop level. We show that 
they are $L$-independent.
\par We set the massless framework in Section IV, using normalization 
conditions as well as minimal subtraction. We discuss the behavior of the 
integral for certain values of $L$ and compute them in the form suitable for 
each renormalization scheme. The exponents obtained from the setting of 
massless fields are computed in Section V. For infinite values of $L$, 
we show that our results correspond to region b). Finite values of $L$ 
are shown to be equivalent to region c). The complete equivalence with 
the exponents computed using massive fields is established. 
\par In Section VI we discuss our results and point out future potential 
applications of the method to approaching other types of critical behaviors 
with simple boundary conditions. Higher loop Feynman integrals are presented 
in the appendixes. The massive integrals in normalization conditions are 
described in Appendix A. In Appendix B we display the massless integrals 
in normalization conditions and in minimal subtraction.      
 
\section{Massive fields for $PBC$ and $ABC$ in the NF Approach} 
\par In this section we begin with a quick review of the $NF$ setup 
\cite{NF2} in order to describe our computation of the critical exponents 
explicitly for periodic and antiperiodic boundary conditions. These 
boundary conditions realize the simplest situation in the discussion of finite 
size effects inasmuch they do not include the effect of free surfaces.  
\par The layered system can be described by the bare Lagrangian density 
\begin{equation}\label{1}
L = \frac{1}{2}
|\bigtriangledown \phi_{0}|^{2} + \frac{1}{2} \mu_{0}^{2}\phi_0^{2} + \frac{1}{4!}\lambda_0\phi_0^{4} ,
\end{equation}
where $\phi_{0}$, $\mu_{0}$ and $g_{0}$ are the bare order parameter, mass 
(where $\mu_{0}^{2}= t_{0}$ is the bare reduced temperature) and coupling 
constant, respectively \cite{amit,BLZ1,BLZ2}. The coordinates 
are decomposed in the form $x=(\vec{\rho},z)$ where $\vec{\rho}$ is a 
$(d-1)$-dimensional vector characterizing the surface of each plate and the 
$z$ direction corresponds to the region perpendicular to them. The 
plates are parallel and layered in the region between  $z=0$ and $z=L$. 
The field satisfies $\phi_{0}(z=0)= \phi_{0}(z=L)$ for periodic 
boundary conditions, whereas $\phi_{0}(z=0)=- \phi_{0}(z=L)$ for antiperiodic 
boundary conditions. The order parameter can be expanded in Fourier modes as 
$\phi_{0}(\vec{\rho},z)= \overset{\infty}{\underset{j=-\infty}{\sum}} \int d^{d-1}k exp(i\vec{k}.\vec{\rho}) 
u_{j}(z) \phi_{0j}(\vec{k})$, where $\vec{k}$ is the momentum vector 
associated to the $(d-1)$-dimensional space, $u_{j}(z)$ are the normalized 
eigenfunctions of the operator $\frac{d^{2}}{dz^{2}}$ whose eigenvalues 
$\kappa_{j}$ defined by $-\frac{d^{2} u_{j}(z)}{dz^{2}}= 
\kappa_{j}^{2} u_{j}(z)$ are called the quasi-momentum along the 
$z$-direction. In addition, the eigenfunctions obey the relations 
$\overset{\infty}{\underset{j=-\infty}{\sum}} u_{j}(z)u_{j}(z')= \delta(z-z')$ and 
$\int_{0}^{L} dz u_{j}(z)u^{*}_{j'}(z)= \delta_{j,j'}$. Note that 
$\kappa_{j}= \sigma (j+\tau)$, where 
$\sigma = \frac{2 \pi}{L}$, $j=0, \pm1, \pm2,...,$ the 
label $\tau =0$ corresponds to PBC and $\tau=\frac{1}{2}$ to ABC. The free 
bare massive propagator ($\mu_{0}^{2} \neq 0$) in momentum space for either 
boundary condition is given by the expression 
$G_{0j}^{(\tau)}(k,j) = \frac{1}{k^{2} + \sigma^{2}(j+ \tau)^{2} 
+ \mu_{0}^{2}}$. 
\par Since a typical Feynman integral involves the product 
of many bare propagators $G_{0j}^{(\tau)}$, the Feynman 
rules are modified as follows: beyond the standard tensorial couplings of the 
infinite theory corresponding to a $N$ component order parameter, each momentum line (propagator) must be multiplied by 
$S^{(\tau)}_{j_{1}j_{2}} = \int_{0}^{L} dz u_{j_{1}}(z)u_{j_{2}}(z)$ and the vertices 
are multipled by the tensor $S^{(\tau)}_{j_{1}j_{2}j_{3}j_{4}}= \int_{0}^{L} 
dz u_{j_{1}}(z)u_{j_{2}}(z)u_{j_{3}}(z)u_{j_{4}}(z)$. Furthermore, each 
momentum loop integral in the finite system can be obtained from the 
infinite system through the substitution 
$\int d^{d}k \rightarrow \overset{\infty}{\underset{j=-\infty}{\sum}} \sigma \int d^{d-1}k$. 
The eigenfunctions 
actually depend on $\tau$ and can be written as 
$u_{j}^{(\tau)}(z)= L^{-\frac{1}{2}} exp(i\kappa_{j}z)$ 
which implies $S^{(\tau)}_{j_{1}j_{2}} = \delta_{j_{1}+j_{2}, 0}$ and 
$S^{(\tau)}_{j_{1}j_{2}j_{3}j_{4}}= L^{-1}\delta_{j_{1}+j_{2}+j_{3}+j_{4},0}$. 
This means that the quasi-momentum is ``conserved'' along the $z$ direction 
for periodic and antiperiodic boundary conditions. 
\par Let us define the renormalized $1PI$ vertex parts from the NF 
construction. Although they do depend on the boundary conditions, we shall 
not introduce this additional label on them. Consequently, considering an 
arbitrary $1PI$ divergent (but regularized, say, by a cutoff $\Lambda$) bare 
vertex part including composite operators 
$\Gamma^{(N,M)}$ ($(N,M)\neq(0,2)$), the statement of multiplicative 
renormalizability amounts to finding renormalization functions 
$Z_{\phi}^{(\tau)}, 
Z_{\phi^{2}}^{(\tau)}$ such that the vertex parts defined by 
\begin{equation}\label{2}
\Gamma_{R}^{(N,M)}(p_{l}, i_{l}, Q_{l}, i'_{l}, g, \mu)= (Z_{\phi}^{(\tau)})^{\frac{N}{2}}(Z_{\phi^{2}}^{(\tau)})^{M} \Gamma^{(N,M)}(p_{l}, i_{l}, Q_{l}, i'_{l}, \lambda_{0}, \mu_{0}, \Lambda),
\end{equation} 
are automatically finite (when the regulator $\Lambda$ is taken to infinity). 
\par In the massive framework, the primitive divergent vertex parts of this 
$\lambda \phi^{4}$ field theory are chosen to be renormalized in the standard 
way \cite{amit}, but now they are explicitly dependent on the boundary 
condition, even though we omit the label $\tau$ in all vertex parts. Then, we 
choose the following normalization conditions at zero external momenta 
(and quasi-momenta), namely 
\begin{subequations}\label{3}
\begin{eqnarray}
&& \Gamma_{R}^{(2)}(k=0 , j=0, g, \mu) = \mu^{2} + \sigma^{2} \tau^{2},
\label{3a} \\
&& \frac{\partial\Gamma_{R}^{(2)}(k, j=0, g, \mu )}{\partial k^{2}}|_{k^{2}=0} \
= 1,\label{3b} \\
&& \Gamma_{R}^{(4)}(k_{l}=0, i_{l}=0, g, \mu ) = g  , \label{3c}\\
&& \Gamma_{R}^{(2,1)}(k=0, j=0, Q=0, j'=0, g, \mu) = 1 .\label{3d}
\end{eqnarray}
\end{subequations} 
Note that the normalization condition on the two-point function above amounts 
to choosing the renormalized mass $\mu$ independent of the boundary condition 
\cite{HD}. These conditions are sufficient 
to formulate all vertex parts which can be renormalized multiplicatively.
\par First let us discuss the situation at the critical dimension $d=4$. In 
that case, utilize implicitly a cutoff $\Lambda$ to regularization of the 
integrals and suppose that after the renormalization procedure is defined the 
limit of infinite cutoff can be taken. We can obtain a Callan-Symanzik 
equation which describes the scaling regime through the following steps: 
i) apply the derivative $\frac{\partial}{\partial \mu_{0}^{2}}$ over 
the bare vertex part $\Gamma^{(N,M)}$ ($(N,M)\neq(0,2)$) at fixed 
$\lambda_{0}, \Lambda$ which produces the 
vertex function $\Gamma^{(N,M+1)} (p_{l},i_{l}, Q_{l}, i'_{l}; 0; \lambda_{0}, 
\mu_{0}, \Lambda)$ at zero inserted momentum; ii) rewrite the 
remaining bare vertex parts in terms of the renormalized ones. This results in 
the following expression
\begin{eqnarray}\label{4}
&&(2 \rho \frac{\partial}{\partial \mu^{2}} +
\frac{\alpha}{\mu^{2}} \frac{\partial}{\partial g}
- \frac{1}{2} N \frac{\kappa}{\mu^{2}} - M \frac{\pi}{\mu^{2}})
\Gamma_{R}^{(N,M)} (p_{l}, i_{l}; Q_{l}, i'_{l}, g,
\mu)= \\ \nonumber
&& \Gamma_{R(\tau)}^{(N,M+1)} (p_{l}, i_{l}; Q_{l}, i'_{l},0, g,\mu) ,
\end{eqnarray}  
where $2 \rho 
= \frac{\partial \mu^{2}}{\partial \mu_{0}^{2}} Z_{\phi^{2}}^{(\tau)}$, 
$\frac{\alpha}{\mu^{2}}=  
Z_{\phi^{2}}^{(\tau)} \frac{\partial g}{\partial \mu_{0}^{2}}$, 
$\frac{\kappa}{\mu^{2}}=  Z_{\phi^{2}}^{(\tau)} \frac{\partial lnZ_{\phi}^{(\tau)}}{\partial \mu_{0}^{2}}$, 
$\frac{\pi}{\mu^{2}}=  Z_{\phi^{2}}^{(\tau)} \frac{\partial lnZ_{\phi^{2}}^{(\tau)}}{\partial \mu_{0}^{2}}$. Let the flow functions be defined by the 
expressions
$\beta^{(\tau)}(\mu, g) (= \frac{\alpha}{\rho})
= \mu \frac{\partial g}{\partial \mu}$, 
$\gamma_{\phi}^{(\tau)}(= \frac{\kappa}{\rho})
= \mu \frac{\partial ln Z_{\phi}^{(\tau)}}{\partial \mu}$ and 
$\gamma_{\phi^{2}}^{(\tau)}(= - \frac{\pi}{\rho})
= - \mu \frac{\partial ln Z_{\phi^{2}}^{(\tau)}}{\partial \mu}$. Multiplying 
last equation by $\frac{\mu^{2}}{\rho}$, we obtain its equivalent form in 
terms of these redefinitions as
\begin{eqnarray}\label{5}
&&(\mu \frac{\partial}{\partial \mu} + 
\beta^{(\tau)} \frac{\partial}{\partial g}
- \frac{N}{2} \gamma_{\phi}^{(\tau)} + M \gamma_{\phi^{2}}^{(\tau)})
\Gamma_{R}^{(N,M)} (p_{l}, i_{l}, Q_{l}, i'_{l}, g,
\mu)= \\ \nonumber
&& 2 \mu^{2} \frac{\partial \mu_{0}^{2}}{\partial \mu^{2}} 
[Z_{\phi^{2}}^{(\tau)}]^{-1} \Gamma_{R}^{(N,M+1)} (p_{l}, i_{l}, 
Q_{l}, i'_{l};0, g, \mu)  \;\;.
\end{eqnarray} 
Now, taking $N=2$ at zero external momenta and quasi-momenta and 
using the normalization conditions Eq.(\ref{3a}) and Eq.(\ref{3d}) we obtain the Callan-Symanzik 
equation for finite size systems given by
\begin{eqnarray}\label{6}
&&(\mu \frac{\partial}{\partial \mu} + 
\beta^{(\tau)} \frac{\partial}{\partial g}
- \frac{N}{2} \gamma_{\phi}^{(\tau)} + M \gamma_{\phi^{2}}^{(\tau)})
\Gamma_{R}^{(N,M)} (p_{l}, i_{l}, Q_{l}, i'_{l}, g,
\mu)= \\ \nonumber
&& \mu^{2} (2 - \tilde{\gamma}_{\phi}^{(\tau)})\Gamma_{R}^{(N,M+1)} (p_{l}, i_{l}, 
Q_{l}, i'_{l};0, g, \mu)  \;\;, 
\end{eqnarray}
where $\tilde{\gamma}_{\phi}^{(\tau)}= \gamma_{\phi}^{(\tau)}[1 + 
(\frac{\sigma \tau}{\mu})^{2}]$. The difference with respect to the original 
version for infinite system is the appearance of 
$\tilde{\gamma}_{\phi}^{(\tau)}$ which is essentially a scaled version of 
$\gamma_{\phi}^{(\tau)}$. Note that in the limit $L\rightarrow \infty$ 
($\sigma \rightarrow 0$) we retrieve the original bulk theory naturally within 
this construction. For fixed $L$, all the discussion of inductive proof of 
multiplicative renormalizability follows exactly as in the bulk system 
described by the ordinary $\phi^{4}$ theory. 
\par Away from the critical dimension, the coupling constant has mass 
dimension. As before, let 
$\beta^{(\tau)}(g,\mu)= \mu \frac{\partial g}{\partial \mu}$ 
be the function which governs the flow of the coupling constant in parameter 
space. In order to get rid of undesirable dimensionful parameters when 
$d=4-\epsilon$, define 
the Gell-Mann-Low function $[\beta(g,\mu)]_{GL}= -\epsilon g + \beta(g, \mu)$. 
Using the Gell-Mann-Low function into the $CS$ equation, it is easy to find 
that all dimensionful parameters turn into 
dimensionless quantities. For instance, let $\lambda= \mu^{\epsilon}u_{0}$ 
be the dimensionful bare coupling constant written in terms of the bare 
dimensionless coupling $u_{0}$ and $g$ the renormalized dimensionful 
counterpart written  in terms of the dimensionless  renormalized 
coupling $u$ as $g= \mu^{\epsilon}u$. Those definitions imply that 
$[\beta(g,\mu)]_{GL} \frac{\partial}{\partial g}= \beta(u) \frac{\partial}
{\partial u}$, i.e., we get a description entirely in terms of the 
dimensionless renormalized coupling constant, which has a well defined scaling 
limit \cite{Vladimirov,Naud}. The Callan-Symanzik equation can be expressed 
in the form
\begin{eqnarray}\label{7}
&&(\mu \frac{\partial}{\partial \mu} + 
\beta(u) \frac{\partial}{\partial u}
- \frac{N}{2} \gamma_{\phi}^{(\tau)} + M \gamma_{\phi^{2}}^{(\tau)})
\Gamma_{R}^{(N,L)} (p_{l}, i_{l}, Q_{l}, i'_{l}, u, \mu)= \\ \nonumber
&& \mu^{2}(2 - \tilde{\gamma}_{\phi}^{(\tau)}) \Gamma_{R}^{(N,L+1)} (p_{l}, i_{l}, 
Q_{l}, i'_{l};0, u, \mu)  \;\;,
\end{eqnarray} 
where $\beta^{(\tau)}(u)= -\epsilon (\frac{\partial lnu_{0}^{(\tau)}}{\partial u})$, $\gamma_{\phi}^{(\tau)}(u) = \beta^{(\tau)} 
(\frac{\partial ln Z_{\phi}^{(\tau)}}{\partial u})$ 
and $\gamma_{\phi^{2}}^{(\tau)} = \beta^{(\tau)} 
(\frac{\partial ln Z_{\phi^{2}}^{(\tau)}}{\partial u})$. The definition 
$\bar{Z}_{\phi^{2}}^{(\tau)}= Z_{\phi}^{(\tau)}Z_{\phi}^{(\tau)}$ can be 
used to write down another function, namely 
$\bar{\gamma}_{\phi^{2}}^{(\tau)} = \beta^{(\tau)} 
(\frac{\partial ln \bar{Z}_{\phi^{2}}^{(\tau)}}{\partial u})$ which shall be 
useful to our purposes. The solution of the Callan-Symanzik equation is 
analogous to the infinite systems version and we shall not discuss it here; 
instead, we shall use the results of previous analysis in order to 
discuss the salient features which naturally leads to the ultraviolet fixed 
points along with the critical exponents for finite systems satisfying 
various boundary conditions.      
\par Recalling that the infrared divergences are absent in the massive theory, 
we analyze the theory at the ultraviolet region where the momentum of the 
internal propagators in arbitrary loop graphs are very large, i.e., at the 
scaling region $\frac{p}{\mu} \rightarrow \infty$ \cite{BLZ1,BLZ2}. This 
means that the right hand side can be neglected order by order in perturbation 
theory just like in the field-theoretic description of infinite systems.
\par Let us turn now our attention to the computation of the Feynman integrals 
corresponding to one-, two- and three-loop diagrams required to getting the 
critical exponents $\eta$ and $\nu$ perturbatively.  
\par The one-loop integral 
contributing to the four-point function is then given by:
\begin{eqnarray}\label{8}
I_{2}^{(\tau)} (k, i; \sigma, \mu) &=& \sigma \sum_{j=-\infty}^{\infty} 
\int d^{d-1}q \frac{1}{[(q)^{2} + (\sigma)^{2}(j+\tau)^{2}+ \mu^{2}]} \nonumber
\\  
&& \times \frac{1}{[(q+k)^{2} + (\sigma)^{2}(j+ i+ \tau)^{2} + \mu^{2}]}.
\end{eqnarray}
Remember that $\mu= t^{\frac{1}{2}}= \xi^{-1}$ at tree level, where $t$ is 
the renormalized reduced temperature. Performing the transformation 
$p=\frac{q}{\mu}$ in all momenta present in the diagram ($k'=\frac{k}{\mu}$, 
restoring $k' \rightarrow k$) and defining 
$r \equiv \frac{\sigma}{\mu}=(\frac{2 \pi \xi}{L})$, we use 
a Feynman parameter $x$ before resolving the integral over $p$, or in other 
words, 
\begin{eqnarray}\label{9}
&& I_{2}^{(\tau)} (k, i; \sigma, \mu) = r \mu^{-\epsilon} 
\sum_{j=-\infty}^{\infty} \int_{0}^{1} dx \int d^{d-1}p \nonumber\\
&& \;\; \times\;\;\frac{1}{\Bigl[p^{2} + 2xkp + xk^{2} 
+ r^{2}[(j+\tau + ix)^{2} + x(1-x)i^{2}] + 1 \Bigr]^{2}}.
\end{eqnarray}
A typical result within our conventions 
(see Ref.\cite{amit}) appropriate to dimensionally regularized integrals is 
expressed by the formula
\begin{equation}\label{10}
\int \frac {d^{d}q}{(q^{2} + 2 k.q + m^{2})^{\alpha}} =
\frac{1}{2} \frac{\Gamma(\frac{d}{2}) \Gamma(\alpha - \frac{d}{2}) (m^{2} - k^{2})^{\frac{d}{2} - \alpha}}{\Gamma(\alpha)} S_{d},
\end{equation} 
where $S_{d}$ is the area of the $d$-dimensional unit sphere. Using this 
relation we get to
\begin{eqnarray}\label{11}
I_{2}^{(\tau)} (k, i; \sigma, \mu) &=& r \mu^{-\epsilon} 
\frac{1}{2} S_{d-1} \Gamma(\frac{d-1}{2}) \Gamma(2 - \frac{(d-1)}{2}) \nonumber \\
&& \times \int_{0}^{1} dx \sum_{j=-\infty}^{\infty} 
[x(1-x) (k^{2} + i^{2} r^{2})+ r^{2} (j + \tau + ix)^{2} + 1]^{\frac{d-1}{2} - 2}.
\end{eqnarray}
\par Notice that 
$r^{-1} \propto \frac{L}{\xi}$ here is the boundedness variable in the massive 
theory, where $\xi$ is the fixed bulk correlation length. After factoring out 
the $r^{2}$ term in the last integral, we can proceed in 
the computation by noticing that the remaining summation can be identified 
with the generalized thermal function \cite{Farina}
\begin{eqnarray}\label{12}
&D_{\alpha}(a,b)= \overset{\infty}{\underset{n=-\infty}{\sum}} [(n+a)^{2} + b^{2}]^{-\alpha} 
\nonumber\\
& = \frac{\sqrt{\pi}}{\Gamma(\alpha)}\;[\frac{\Gamma(\alpha - \frac{1}{2})}{b^{2\alpha -1}} + f_{\alpha}(a,b)], 
\end{eqnarray}
where 
\begin{equation}\label{13} 
f_{\alpha}(a,b)= 4 \sum_{m=1}^{\infty} cos(2\pi ma)
(\frac{\pi m}{b})^{\alpha - \frac{1}{2}} K_{\alpha - \frac{1}{2}}(2\pi mb),
\end{equation} 
and $K_{\nu}(x)$ is the modified Bessel function of the second kind. The 
identifications $a(x)=\tau + ix$, $b(x)=r^{-1}\sqrt{(k^{2} + r^{2})x(1-x) +1}$ 
and $\epsilon=4-d$ permit us to write
\begin{eqnarray}\label{14}
I_{2}^{(\tau)} (k, i; \sigma, \mu) &=& \mu^{-\epsilon} 
\frac{1}{2} S_{d-1} \Gamma(\frac{d-1}{2}) \sqrt{\pi}\; [\int_{0}^{1} dx \nonumber 
\\
&& \times \;\;[\Gamma(\frac{\epsilon}{2}) 
[x(1-x) (k^{2} + i^{2} r^{2})+ 1]^{-\frac{\epsilon}{2}} 
+ f_{\frac{1}{2} + \frac{\epsilon}{2}}(a,b)]], 
\end{eqnarray}
Now, using the identity 
$\sqrt{\pi}\Gamma(\frac{d-1}{2})S_{d-1}= \Gamma(\frac{d}{2}) S_{d}$ 
and expanding in $\epsilon$ the argument of the gamma function, we can 
rewrite last integral as
\begin{eqnarray}\label{15}
I_{2}^{(\tau)} (k, i; \sigma, \mu) &=& S_{d}\mu^{-\epsilon} [
\frac{1}{\epsilon}(1 - \frac{\epsilon}{2})
\times\;\; \int_{0}^{1} dx [x(1-x) (k^{2} + i^{2} r^{2})+ 1]^{-\frac{\epsilon}{2}} \nonumber\\
&&  +  \frac{1}{2} r^{-\epsilon} \Gamma(2-\frac{\epsilon}{2})\int_{0}^{1} dx f_{\frac{1}{2} + \frac{\epsilon}{2}}(\tau + ix, r^{-1}\sqrt{x(1-x) (k^{2} + i^{2} r^{2})+ 1})]. 
\end{eqnarray}
\par Whenever 
we perform a loop integral, the area of the unit sphere $S_{d}$ naturally 
takes place and this angular factor can be neutralized in a redefinition of 
the coupling constant. We adopt this procedure henceforward in all loop 
integrals and suppress this overall factor. We then find 
\begin{eqnarray}\label{16}
& I_{2}^{(\tau)} (k,i;r) \equiv \frac{I_{2}^{(\tau)} (k,i;\sigma, \mu)}{S_{d}}= \frac{\mu^{-\epsilon}}{\epsilon}
\bigl((1-\frac{\epsilon}{2})\int_{0}^{1} dx [x(1-x)(k^{2} 
\nonumber\\
& + r^{2} i^{2}) + 1]^{ - \frac{\epsilon}{2}} + \frac{\epsilon}{2} \Gamma(2-\frac{\epsilon}{2}) F^{(\tau)}_{\frac{\epsilon}{2}}(k,i;r)\bigr),
\end{eqnarray} 
where 
\begin{equation}\label{17}
F^{(\tau)}_{\alpha}(k,i;r)= r^{-2 \alpha} \int_{0}^{1} dx 
f_{\frac{1}{2} + \alpha}\Bigl(\tau+xi, h(k, i, r) \Bigr), 
\end{equation}
and 
\begin{equation}\label{18} 
h(k, i, r)= r^{-1} \sqrt{x(1-x)(k^{2} + r^{2} i^{2}) +1}.
\end{equation}
Since we are going to use normalization conditions 
in this massive setting, we are interested in the simplest situation 
which occurs for vanishing external momenta and quasi-momenta $(k=0, i=0)$. 
In that case, $F^{(\tau)}_{\frac{\epsilon}{2}}(r)= r^{-\epsilon}
f_{\frac{1}{2} + \frac{\epsilon}{2} }(\tau, r^{-1})$. Furthermore, recalling 
that the finite size contribution is $O(\epsilon^{0})$, the one-loop integral 
can be written as 
\begin{eqnarray}\label{19}
I_{2}^{(\tau)} (r) \equiv I_{2}^{(\tau)}(k=0,i=0,r)= \mu^{-\epsilon} 
\bigl[\frac{1}{\epsilon}
(1-\frac{\epsilon}{2}) + \frac{1}{2} f_{\frac{1}{2}}(\tau,r^{-1})\bigr].
\end{eqnarray} 
In this massive approach, so long as $r^{-1}$ takes finite nonzero values, 
last equation is suitable to compute amplitudes and other observables and 
demonstrates explicitly that the one-loop bubble at the symmetry point is 
decomposable in the form bulk contribution plus finite size 
correction. 
\par On the other hand, in the computation of the critical exponents, 
in practice we use this form without any specification of the correction 
function, since it does not have singular behavior for these values of 
$r^{-1}$ and the divergence structure of this diagram is governed by the 
first term (bulk) in that expression. In that case, the $\epsilon$-expansion 
is well defined and we can proceed to the computation of loop diagrams of 
primitively divergent vertex parts in order to renormalize the theory and 
obtain those universal quantities. Note that even though 
the last expression is written in terms of $f_{\frac{1}{2}}(\tau, r^{-1})$, 
we prefer to write the correction in terms of 
$F^{(\tau)}_{\frac{\epsilon}{2}}(k=0,i=0;r)$ anticipating our future 
discussion of the massless case. In addition, applications of the present 
method might be important to compute amplitudes. We would like to 
understand the importance of the finite size correction in limit values of the 
boundedness variable in calculating an arbitrary amplitude. This amounts to 
figuring out the approach to the regions $r^{-1} \rightarrow 0$ 
(or $r \rightarrow \infty$) as well as $r^{-1} \rightarrow \infty$ and what 
are the limits of validity of the $\epsilon$-expansion.
\par In order to describe these asymptotic values and its effects on the 
finite size correction, we write the latter in the form
\begin{equation}\label{20}
F^{(\tau)}_{\frac{\epsilon}{2}}(k=0,i=0;r) = 4 r^{-\epsilon} 
\sum_{n=1}^{\infty} 
cos(2 \pi n\tau)(\pi n)^{\frac{\epsilon}{2}}
K_{\frac{\epsilon}{2}}(2 \pi n r^{-1}).
\end{equation} 
\par This expression shows clearly that the correction has no poles in 
$\epsilon$. Since it is well behaved, take $\epsilon=0$
in this whole expression in order to rewrite Eq.(\ref{19}) at the 
symmetry point as 
\begin{equation}\label{21}
I_{2SP}^{(\tau)} (r)= \mu^{-\epsilon} \bigl[\frac{1}{\epsilon}
\bigl(1-\frac{\epsilon}{2}\bigr) +  2  \sum_{n=1}^{\infty} 
cos(2 \pi n\tau)K_{0}(2 \pi n r^{-1})\bigr]. 
\end{equation} 
\par The limit $r^{-1} \rightarrow \infty$ corresponds to 
$\frac{L}{\xi} \rightarrow \infty$, whereas $0< r^{-1}< \infty$ represent finite values of $\frac{L}{\xi}$. Let us focus our attention in the limit $r^{-1}\rightarrow \infty$. Using 
the asymptotic form of the Bessel function for $x \rightarrow \infty$, namely 
$K_{\alpha}(x) = \sqrt{\frac{\pi}{2x}} e^{-x}[1 + O(1/x)]$, for 
$r^{-1}\rightarrow \infty$ one learns that the correction term has the behavior
\begin{equation}\label{22}
\underset{r^{-1} \rightarrow \infty}{lim} \bigl(\frac{\epsilon}{2}
F^{\tau}_{\frac{\epsilon}{2}}(k=0,i=0;r)\bigr) = \frac{2\epsilon}{(r^{-1})^{\frac{1}{2}}} \sum_{n=1}^{\infty} cos(2 \pi n\tau) n^{-1/2} e^{(-2 \pi n r^{-1})}.
\end{equation}
Using a trivial unequality to simplify our task, we can show that last term 
vanishes in the wanted limit as follows
\begin{eqnarray}\label{23}
&& \underset{r^{-1} \rightarrow \infty}{lim} \bigl(\frac{1}{r^{-\frac{1}{2}}} 
\sum_{n=1}^{\infty} cos(2 \pi n\tau) n^{-1/2} e^{(-2 \pi n r^{-1})}\bigr) < 
\underset{r^{-1} \rightarrow \infty}{lim} \bigl(\frac{1}{r^{-\frac{1}{2}}} 
\sum_{n=1}^{\infty} cos(2 \pi n\tau) e^{(-2 \pi n r^{-1})}\bigr) = \nonumber\\
&& \underset{r^{-1} \rightarrow \infty}{lim} 
\bigl[\frac{1}{r^{-\frac{1}{2}}}\bigr] \bigl(\frac{1}{1-e^{-2\pi (r^{-1} -i\tau)}} 
+ \frac{1}{1-e^{-2\pi (r^{-1} +i\tau)}} -2\bigr) \rightarrow 0.
\end{eqnarray}
Therefore, the integral turns out to reproduce the (bulk) value from the 
massive theory of the infinite system \cite{BLZ1}. We can identify 
region $\frac{L}{\xi} \rightarrow \infty$ with usual bulk critical behavior 
$L \rightarrow \infty$. The region $\frac{L}{\xi}>1$ interpolates from  
finite size corrections to the bulk critical behavior. 
\par As $\frac{L}{\xi}$ decreases, the finite size 
correction will increase until it will eventually become as big as the pole 
in $\epsilon$, modifying the leading singularity of the four-point function. To see this let us consider the potential trouble which is hidden in the 
different values of $r^{-1}$ and, in particular, in the limit 
$r^{-1} \rightarrow 0$ ($L \rightarrow 0$).
\par Let us perform the sum which appears explicitly in the correction term. 
From Ref.\cite{GR}, the identity 
\begin{eqnarray}\label{24}
\sum_{n=1}^{\infty} K_{0}(n x) cos(n x t) =&& \frac{1}{2} [\gamma
+ ln(\frac{x}{4 \pi})] + \frac{\pi}{2x \sqrt{1+t^{2}}} + \frac{\pi}{2} 
\sum_{n=1}^{\infty} 
\bigl[\frac{1}{\sqrt{x^{2} + (2n\pi + tx)^{2}}} \nonumber\\
&& - \frac{1}{2 n \pi}\bigr] + \frac{\pi}{2} \sum_{n=1}^{\infty} 
\bigl[\frac{1}{\sqrt{x^{2} + (2n\pi - tx)^{2}}} - \frac{1}{2 n \pi}\bigr], 
\end{eqnarray}
which is valid for positive finite values of the variable $x$, 
along with the identifications $x= 2 \pi r^{-1}$, $t= r \tau$ (and 
$\gamma = 0.57721566...$ is the Euler-Mascheroni constant) 
implies the following result to the one-loop graph
\begin{eqnarray}\label{25}
&& I_{2}(k=0, i=0, r^{-1})= \mu^{-\epsilon} \bigl[\frac{1}{\epsilon}
(1-\frac{\epsilon}{2}) + \gamma + ln(\frac{r^{-1}}{2}) + \frac{1}{2 \sqrt{r^{-2} + \tau^{2}}} \nonumber\\
&& + \frac{1}{2}\sum_{n=1}^{\infty} 
\bigl[\frac{1}{\sqrt{r^{-2} + (n + \tau)^{2}}} - \frac{1}{n}\bigr] 
+ \frac{1}{2}\sum_{n=1}^{\infty} 
\bigl[\frac{1}{\sqrt{r^{-2} + (n - \tau)^{2}}} - \frac{1}{n}\bigr] \bigr]. 
\end{eqnarray} 
\par The simplest way to prove that the two infinite series are convergent 
in the limit $r^{-1}\rightarrow 0$ is to set directly $r^{-1}=0$ and compute 
this correction \cite{RGJ}. It becomes 
\begin{eqnarray}\label{26}
\underset{r^{-1} \rightarrow 0}{lim}\bigl(\frac{1}{2}\sum_{n=1}^{\infty} 
\bigl[\;\;\bigr] 
+ \frac{1}{2}\sum_{n=1}^{\infty} 
\bigl[\;\;\bigr]\bigr)= \frac{1}{2}\sum_{n=1}^{\infty} 
\bigl[\frac{1}{n + \tau} - \frac{1}{n}\bigr] 
+ \frac{1}{2}\sum_{n=1}^{\infty} 
\bigl[\frac{1}{n - \tau} - \frac{1}{n}\bigr]. 
\end{eqnarray} 
Now, from the definition of the dilogarithm function 
$\psi(1+z)= -\gamma + \sum_{n=1}^{\infty} \frac{z}{(n + z)}$, together 
with the relation $\psi(1+z)= \psi(z) + \frac{1}{z}$ and the value 
$\psi(\frac{1}{2})= -\gamma -2ln2$, we easily obtain
\begin{eqnarray}\label{27}
\underset{r^{-1} \rightarrow 0}{lim}\bigl(\frac{1}{2}\sum_{n=1}^{\infty} 
\bigl[\;\;\bigr] 
+ \frac{1}{2}\sum_{n=1}^{\infty} 
\bigl[\;\;\bigr]\bigr)= (2ln2-1)\delta_{\tau,\frac{1}{2}},
\end{eqnarray} 
which is finite as advertised. Therefore, for small values of $r^{-1}$ we 
can write the one-loop bubble as 
\begin{eqnarray}\label{28}
&& I_{2}(k=0, i=0, r^{-1})= \mu^{-\epsilon}\bigl[\frac{1}{\epsilon}
(1-\frac{\epsilon}{2}) + \gamma + ln(\frac{r^{-1}}{2}) + \frac{1}{2 \sqrt{r^{-2} + \tau^{2}}} \nonumber\\
&& + (2ln2-1) \delta_{\tau,\frac{1}{2}} + O(r^{-1})\bigr]. 
\end{eqnarray} 
 \par This expression for the one-loop four point function depends on the 
boundary condition. It gives the support to identify two types of 
crossover in finite systems presenting these simple boundary conditions 
away from the critical point ($t\neq0$) as follows.
\par Firstly, our analytical expression above is a transliteration of the 
analysis performed in Refs.\cite{NF1,NF2} concerning the breakdown of the 
expansion in $\epsilon=4-d$, namely, when the argument of the square root 
term in the above expression vanishes. Indeed, for periodic boundary 
conditions $\tau=0$ and perturbation theory is invalid in the limit 
$r^{-1} = \frac{L}{2\pi \xi} \rightarrow 0$. For antiperiodic boundary 
conditions, however, if the temperature is below the bulk critical 
temperature ($t<0$), whenever 
$r^{-2}= -\frac{L^{2}}{(2\pi \xi)^{2}}=\frac{1}{2}$ the inverse square root 
blows up. 
This effect was denominated ``dimensional crossover'' as discussed previously 
by those authors.  
\par Secondly, if the value of $r^{-1}$ is decreased further for fixed 
$t>0$, i.e., diminishing $L$, the logarithm term starts to become important 
for antiperiodic boundary condition when its argument becomes around the 
same order of magnitude that the dimensional pole $\frac{1}{\epsilon}$. 
If we switch to cutoff regularization for a moment, 
the ultraviolet regime is characterized by $\frac{1}{\epsilon} \rightarrow 
ln(\frac{\Lambda}{\mu})= ln(\Lambda \xi)$ with $\Lambda \xi \gg 1$. The 
logarithm contribution will eventually become comparable with the ultraviolet 
dimensional pole, whenever 
$(\frac{L}{\xi}) \sim \frac{1}{(\Lambda \xi)}$, i.e., 
when $L\Lambda \sim 1$. In terms of a lattice parameter $a$, 
$\Lambda \sim \frac{1}{a}$ which implies $L \sim a$. It is the reduction of 
$L$ for fixed $t>0$ in this massive framework which is responsible for 
this new effect. This is a novel type 
of crossover which only happens for antiperiodic boundary condition at $t>0$ 
and is straightforward from our analytical expression given purely in terms 
of elementary primitive functions. This new type of crossover starts when 
instead of a large number of parallel plates, there are only two parallel 
plates (the limiting surfaces) and the bulk description is no longer reliable. 
Note that this behavior is also there for periodic boundary condition, but 
the square root term proportional to $\bigl(\frac{L}{\xi}\bigr)^{-1}$ 
is overwhelming in that limit.  
\par A word of caution here. It is dangerous to take the limit $t=0$ (or 
$\xi \rightarrow \infty$) in the above expression. The reason this limit is 
inconsistent in this massive framework is that the scale invariance of the 
renormalized theory only takes place in the ultraviolet regime. The most 
appropriate strategy would be to start from scratch with massless fields which 
are scale invariant at this infrared regime, renormalize the theory at 
nonvanishing external momenta scale and push forward all the consequences 
which follow from this approach. We are going to study this case later on and 
shall prove from a full two-loop calculation that the phenomenological 
scaling theory, which states that $\epsilon$-expansion results have 
meaningless results at $t=0$ is incorrect. We postpone this discussion to 
Secs. IV and V.  
\par Without loss of generality we 
can choose $\mu^{2}=1$ which is 
equivalent to a fixed (arbitrary but finite) correlation length, 
such that $r=\sigma= \frac{2\pi}{L}$. (In our subsequent discussion we 
can reconstruct the $\xi$ dependence through its multiplication by $L^{-1}$.) 
In fact we could have started directly 
from this choice for the mass scale, and it will define all other massive loop 
integrals yet to be discussed. We have only to keep in mind that this choice 
makes $L$ dimensionless. 
\par From now on, we stay away from the region of crossover in 
order to compute the higher loop integrals. These objects can be computed 
analogously to our previous one-loop discussion and the reader is 
advised to consult Appendix A for details. One typical example is the 
integral contributing to the four-point function at two-loops, namely 
\begin{eqnarray}\label{29}
&&I_{4}(k_{1},k_{2},k_{3}, k_{4},i_{1},i_{2},i_{3}, i_{4},\sigma,\mu) = \sigma^{2}\sum_{j_{1},j_{2}=-\infty}^{\infty}
\int \frac{d^{d-1}{q_{1}}d^{d-1}q_{2}}
{\left(q_{1}^{2} + \sigma^{2}(j_{1} + \tau)^2 + \mu^{2} \right)} \nonumber\\
&& \;\times \frac{1}{\left(q_{2}^{2} + \sigma^{2}(j_{2} + \tau)^2 + \mu^{2} \right)[(q_{1} - q_{2} + k_{3})^{2} + \sigma^{2}(j_{1} - j_{2} + i_{3} + \tau)^2
+ \mu^{2}]} \nonumber\\
&&\;\;\times \frac{1}{[(P - q_{1})^{2} + \sigma^{2}(p - j_{1} 
+ \tau)^2 + \mu^{2}]}\;\;,
\end{eqnarray}
where $P=k_{1}+k_{2}$ is the external momenta along the plates and 
$p=i_{1}+i_{2}$ is a discrete ``external'' quasi-momentum label. Taking 
$\mu=1$ at zero external momenta and quasi-momenta, let the integral be denoted by $I_{4}^{(\tau)}(0,0;\sigma)$. From the result 
computed in Appendix A, the outcome to this diagram is
\begin{equation}\label{30}
I_{4SP}^{(\tau)} (\sigma)= \frac{1}{2 \epsilon^{2}}
\bigl((1-\frac{\epsilon}{2}) + \epsilon f_{\frac{1}{2}}(\tau, \sigma^{-1})\bigr).
\end{equation}
\par The integrals contributing to the two-point function at two- and 
three-loops, respectively, turn out to be
\begin{eqnarray}\label{31}
&& I_{3}(k,i,\sigma,\mu) = \sigma^{2}\sum_{j_{1},j_{2}=-\infty}^{\infty} 
\int \frac{d^{d-1}{q_{1}}d^{d-1}q_{2}}
{\left( q_{1}^{2} + \sigma^{2}(j_{1}+\tau)^2 + \mu^{2} \right)
\left( q_{2}^{2} + \sigma^{2}(j_{2}+\tau)^2 + \mu^{2} \right)} \nonumber\\
&& \times \;\;\; \frac{1}{[(q_{1} + q_{2} + k)^{2} 
+ \sigma^{2}(j_{1} + j_{2}+ i + \tau)^{2} + \mu^{2}]}\;\;,
\end{eqnarray}
and
\begin{eqnarray}\label{32}
&& I_{5}(k,i,\sigma,\mu) = \sigma^{2}\sum_{j_{1},j_{2}=-\infty}^{\infty}
\int \frac{d^{d-1}{q_{1}}d^{d-1}q_{2}d^{d-1}q_{3}}
{\left(q_{1}^{2} + \sigma^{2}(j_{1}+\tau)^{2} + \mu^{2} \right) 
\left( q_{2}^{2} +  \sigma^{2}(j_{2}+\tau)^{2} 
+ \mu^{2} \right)} \nonumber\\
&& \times \frac{1}{\left( q_{3}^{2} +  \sigma^{2}(j_{3}+\tau)^{2} 
+ \mu^{2} \right)[ (q_{1} + q_{2} + k)^{2} 
+ \sigma^{2}(j_{1}+j_{2}+i+\tau)^{2} + \mu^{2}]} \nonumber\\
&& \times \frac{1}{[(q_{1} + q_{3} + k)^{2} 
+ \sigma^{2}(j_{1}+j_{3}+i+\tau)^{2} + \mu^{2}]}.
\end{eqnarray}
In passing, we note that the massless diagrams to be studied later, 
can be obtained from the above expressions by setting $\mu=0$, although we 
work in the massive case at $\mu=1$. 
\par The objects required for our purposes are the derivative of those 
integrals with respect to the external momenta, setting the external momenta 
at the symmetry point in the end of the process. Let the derivatives in 
relation to $k^{2}$ be $I_{3}^{\prime (\tau)}$ and $I_{5}^{\prime (\tau)}$ 
(at null $k$). It is demonstrated in Appendix A that they are 
given by:
\begin{equation}\label{33}
I_{3SP}^{\prime (\tau)} (\sigma)= -\frac{1}{8\epsilon}
\bigl(1-\frac{\epsilon}{4} + \epsilon W^{(\tau)}(\sigma)\bigr), 
\end{equation}
and 
\begin{equation}\label{34}
I_{5SP}^{\prime (\tau)} (\sigma)= -\frac{1}{6\epsilon^{2}}
\bigl(1-\frac{\epsilon}{4} + \frac{3 \epsilon}{2} W^{(\tau)}(\sigma)\bigr),
\end{equation} 
where $W^{(\tau)}(\sigma)= G^{(\tau)}(\sigma) + H^{(\tau)}(\sigma) - 
4 r F_{0}^{\prime (\tau)}(\sigma)$ and $F_{0}^{\prime (\tau)}(\sigma)$, 
$G^{(\tau)}(\sigma)$ and $H^{(\tau)}(\sigma)$ are defined in Appendix A, 
Eqs.(\ref{A7}), (\ref{A14a}) and (\ref{A14b}), respectively, computed at 
$\epsilon=0$. We are not going to write them down explicitly since we shall 
show soon that they will be eliminated during the renormalization process. 
\par With the information at hand, we can proceed to compute the 
normalization functions, Wilson functions, the repulsive fixed point, the 
anomalous dimension of the field and of the composite operator. This task 
shall be tackled in the next section.

\section{Critical exponents from finite size with PBC and ABC}
\par Now we describe the normalization functions and Wilson functions in 
terms of the loop integrals. These are the fundamental quantities needed 
to uncover the diagrammatic computation of universal quantities. 
\par The occurrence of a nontrivial ultraviolet fixed point, the scaling 
limit in the ultraviolet regime and the simplification achieved for the 
renormalized vertex parts computed at the fixed point are important aspects 
with consequences in this sort of computation. 
\par As previously discussed, the ultraviolet flow in 
momentum space can be described in terms of dimensionless coupling constants. 
In this way, we can write the dimensionless bare coupling constant and 
normalization function $Z_{\phi}$, $\bar{Z}_{\phi^{2}}$ as power 
series in the dimensionless renormalized coupling constant as 
$u_{0}^{(\tau)} = u (1 + a_{1}^{(\tau)}u + a_{2}^{(\tau)}u^{2})$, 
$Z_{\phi}^{(\tau)} = 1 + b_{2}^{(\tau)} u^{2} + b_{3}^{(\tau)} u^{3}$ and 
$\bar{Z}_{\phi^{2}}^{(\tau)} = 1 + c_{1}^{(\tau)} u + c_{2}^{(\tau)} u^{2}$. 
The divergence structure of these objects are dimensional poles appearing 
as inverse powers of $\epsilon(=4-d)$. In order to figure out explicitly each 
coefficient as parameters depending on the loop Feynman integrals 
computed (at symmetry point defined at zero external momenta and 
quasimomenta) so far, express them in the form
\begin{subequations}\label{35}
\begin{eqnarray}
u_{0} =&& u [1 + \frac{(N+8)}{6} I_{2 SP}^{(\tau)} u 
+ (\frac{[(N+8)I_{2 SP}^{(\tau)}]^{2}}{18}\nonumber\\ 
&& - (\frac{(N^{2}+6N+20)(I_{2 SP}^{(\tau)})^{2}}{36} + 
\frac{(5N+22)I_{4SP}^{(\tau)}}{9}) - \frac{(N+2)I_{3SP}^{' (\tau)}}{9})u^{2}],\\
Z_{\phi}^{(\tau)} =&& 1 + \frac{(N+2)I_{3 SP}^{' (\tau)}}{18} u
+ \frac{(N+2)(N+8)(I_{2SP}^{(\tau)} I_{3 SP}^{' (\tau)} - \frac{I_{5SP}^{' (\tau)}}{2})}{54}u^{2},\\
\bar{Z}_{\phi^{2}}^{(\tau)} =&& 1 + \frac{(N+2)I_{2SP}^{(\tau)}}{6} u \nonumber \\
&& + [\frac{(N^{2}+7N+10)I_{2SP}^{(\tau)}}{18} - \frac{(N+2)}{6}(\frac{(N+2)
(I_{2SP}^{(\tau)})^{2}}{6} 
+ I_{4SP}^{(\tau)})]u^{2}.
\end{eqnarray}
\end{subequations}
\par The flow functions describing the parameter space as the momentum scale 
varies are $\beta^{(\tau)}(u)$, $\gamma_{\phi}^{(\tau)}(u)$ 
and $\bar{\gamma}_{\phi^{2}}^{(\tau)}$. When they are written as 
series expansions in terms of $u$, we obtain explicitly 
\begin{subequations}\label{36}
\begin{eqnarray}
&& \beta^{(\tau)}(u)  =  -\epsilon u[1 - a_{1}^{(\tau)} u
+2((a_{1}^{(\tau)})^{2} - a_{2}^{(\tau)}) u^{2}],\label{36a}\\
&& \gamma_{\phi}^{(\tau)} = -\epsilon u [2b_{2}^{(\tau)} u
+ (3 b_{3}^{(\tau)}  - 2 b_{2}^{(\tau)} a_{1}^{(\tau)}) u^{2}],\label{36b}\\
&& \bar{\gamma}_{\phi^{2}}^{(\tau)} = \epsilon u [c_{1}^{(\tau)}
+ (2 c_{2}^{(\tau)}  - (c_{1}^{(\tau)})^{2} 
- a_{1}^{(\tau)} c_{1}^{(\tau)})u]. \label{36c}
\end{eqnarray}
\end{subequations}
\par We then employ the results for the integrals presented in the last 
section for periodic and antiperiodic boundary conditions in order to find 
the values of each coeficcient. 
\par Looking at Eq.(\ref{36}) and comparing it 
with the expansions of the dimensionless bare coupling contant, normalization 
constant of the field and that from the composite operator, we can read off 
their results. The fact of the matter is that the utilization of normalization 
conditions provokes the appearance of correction terms in those functions 
which depend explicitly on the boundary conditions. As is well known, this 
is a prevalent artifact taking place in this renormalization scheme. Thus, 
the renormalization functions at two loops will not be equal to those 
provenient from the bulk system. Nevertheless, these nonuniversal corrections 
are going to cancel out in the expression for universal quantities as we shall 
see in the remainder of this section. 
\par Working out the details using this prescription, it is not 
difficult to prove that they are given by the following expressions
\begin{subequations}\label{37}
\begin{eqnarray}
&& a_{1}^{(\tau)} = \frac{(N+8)}{6 \epsilon}[1 -\frac{1}{2}\epsilon + \frac{1}{2} f_{\frac{1}{2}}(\tau, \sigma^{-1}) \epsilon] ,\\
&& a_{2}^{(\tau)} = (\frac{N+8}{6 \epsilon})^{2}
- \frac{N^{2}+21N+86}{36 \epsilon} 
+ (\frac{N^{2}+16N+64}{36\epsilon})f_{\frac{1}{2}}(\tau, \sigma^{-1})\nonumber\\
&& \;\;\;\;\;\;\;\;\; + \;\;(\frac{N+2}{72\epsilon}) (1 - \frac{\epsilon}{4} + \epsilon W^{(\tau)}(\sigma)) ,\\
&& b_{2}^{(\tau)} = -\frac{(N+2)}{144 \epsilon}[1 - \frac{\epsilon}{4} 
+ \epsilon W^{(\tau}(\sigma)], \\
&& b_{3}^{(\tau)} = -\frac{(N+2)(N+8)}{1296 \epsilon^{2}}[1 
- \frac{7\epsilon}{4} + \frac{3\epsilon}{2}f_{\frac{1}{2}}(\tau, \sigma^{-1})], \\
&& c_{1}^{(\tau)} = \frac{(N+2)}{6 \epsilon}[1 -\frac{1}{2}\epsilon + \frac{1}{2} f_{\frac{1}{2}}(\tau, \sigma^{-1}) \epsilon], \\
&& c_{2}^{(\tau)} = \frac{(N+2)(N+5)}{36 \epsilon^{2}}
- \frac{2N^{2}+17N+26}{72 \epsilon} 
+ \frac{(N^{2} + 7N +10)}{36 \epsilon} f_{\frac{1}{2}}(\tau, \sigma^{-1}).
\end{eqnarray}
\end{subequations} 
\par Using the coefficients $a_{1}^{(\tau)}$ 
as well as $a_{2}^{(\tau)}$ into Eq.(\ref{36a}) we get to  
\begin{equation}\label{38}
\beta^{(\tau)}(u) = -\epsilon u  + \frac{(N+8)}{6}\bigl(1 - \frac{\epsilon}{2} 
+ \frac{\epsilon}{2} f_{\frac{1}{2}}(\tau, \sigma^{-1})\bigr) u^{2} 
- \frac{3N+14}{12}u^{3}.
\end{equation} 
Furthermore, we can also find the solutions for the functions 
related to the anomalous dimension of the field and composite operator. They 
read
\begin{subequations}\label{39}
\begin{eqnarray}
&& \gamma_{\phi}^{(\tau)} = u^{2} \frac{(N+2)}{72}[1 - \frac{\epsilon}{4} 
+ \epsilon W^{(\tau)}(\sigma) - \frac{(N+8)}{6}(1 + W^{(\tau)}(\sigma) 
- f_{\frac{1}{2}}(\tau, \sigma^{-1}))u],\label{39a}\\
&& \bar{\gamma}_{\phi^{2}}^{(\tau)} = \frac{(N+2)}{6}(1 - \frac{\epsilon}{2} 
+ \frac{\epsilon}{2} f_{\frac{1}{2}}(\tau, \sigma^{-1}))u - \frac{(N+2)}{12} u^{2}.\label{39b}
\end{eqnarray}
\end{subequations}
The eigenvalue condition $\beta^{(\tau)}(u_{\infty})=0$ yields the repulsive 
ultraviolet fixed point which is given by
\begin{equation}\label{40}
u_{\infty}= \frac{6}{(N+8)} \epsilon[1 + \epsilon[\frac{(9N+42)}{(N+8)^{2}} 
+ \frac{1}{2}(1 - f_{\frac{1}{2}}(\tau, \sigma^{-1}))]].
\end{equation} 
As usual we identify the fixed point value of the anomalous dimension of the 
field with the critical exponent $\eta$, i.e.
\begin{equation}\label{41}
\eta=\gamma_{\phi}^{(\tau)}(u_{\infty})= \frac{(N+2)\epsilon^{2}}{(N+8)^{2}}
[1 + (\frac{6(3N+14)}{(N+8)^{2}} - \frac{1}{4})\epsilon].
\end{equation}
Similarly, it is straightforward to show that 
\begin{equation}\label{42}
 \bar{\gamma}_{\phi^{2}}^{(\tau)} (u_{\infty})= \frac{(N+2)\epsilon}{(N+8)} 
[1 + \frac{(6N+18)}{(N+8)^{2}} \epsilon],
\end{equation}  
which in conjuminance with the equation 
$\bar{\gamma}_{\phi^{2}}^{(\tau)} (u_{\infty})= 2 - \eta - \nu^{-1}$ yields 
the value of the correlation lenght exponent
\begin{equation}\label{43}
\nu = \frac{1}{2} + \frac{(N+2)}{4(N+8)} \epsilon 
+ \frac{(N+2)(N^{2}+23N+60)}{8(N+8)^{3}} \epsilon^{2}.
\end{equation}
Note that these exponents are indeed independent of the boundary conditions 
and are exactly the same as those obtained from the bulk system through an 
analogous utilization of diagrammatic methods. If we use the remaining bulk 
scaling relations, we find that all the critical exponents for these simple 
boundary conditions reproduce those from the bulk confirming the               
one loop analysis by Nemirovsky and Freed and extending it to the present higher loop 
correction.
\par We have thus succeeded in formulating $1PI$ renormalized vertex parts for 
the massive theory in order to compute the exponents within the perturbation 
expansion in $\epsilon$. We shall introduce the framework of massless 
fields in the calculation of the critical exponents in the next section.  

\section{NF method for PBC and ABC using massless fields}
\par We start by describing the method of renormalization for massless 
fields, given by the previous Lagrangian density. The $(d-1)$-dimensional 
momentum space lies along the plates and characterizes directions 
perpendicular to the finite size direction $L$,which is represented by the 
quasimomenta.  The free bare critical (massless) propagator is 
$G_{0j}^{(\tau)}(k,j) = \frac{1}{k^{2} + \sigma^{2}(j+ \tau)^{2}}$. 
The former definitions for the tensors $S_{i_{1} i_{2}}$ and 
$S_{i_{1}i_{2}i_{3}i_{4}}$ hold for the massless case as well.
\par Let us focus now in the renormalization scheme to be chosen in the 
massless situation. The massless theory has infrared divergences at zero 
external momenta, so it would be worthy to define the renormalized theory 
of the few bare primitively divergent vertex parts to be finite after 
using the divergent normalization constants. It is very simple to 
employ two independent renormalization schemes in order to compute critical 
exponents: normalization conditions and minimal subtraction. We emphasize 
that the integrals involved can be resolved 
for arbitrary external momenta so that the most convenient form of the results 
can be used to pursue the computation of the critical exponents in either 
renormalization scheme. We postpone this discussion to the next section. For 
the time being we shall analyze basic facts regarding the structure of the 
normalization conditions. In the present section, a thorough analysis of the 
one-loop four-point diagram in different limits shall be worked out along with 
the quotation of higher loop diagrams (extracted from Appendix B).   
\par Normalization conditions is appealing due to its simplicity after the 
choice of the symmetry(subtraction) point for the external momenta taken at 
nonzero value. Let $k_{i}$ be the external momenta of a 
$(d-1)$-dimensional transversal space and let $\kappa$ be the external 
momentum scale along the plates where the renormalized theory is defined. 
At the symmetry point we choose $k_{i} . k_{j}= \frac{\kappa^{2}}{4}
(4\delta_{ij}-1)$ leading to $(k_{i} + k_{j})^{2} = \kappa^{2}$. We fix 
the external momentum scale of the two-point function at $k^{2}= \kappa^{2}=1$.
The multiplicative 
renormalization can be achieved through conditions on the primitively 
divergent bare vertices at zero mass, such that their renormalized versions 
take the following values at the symmetry point:
\begin{subequations}\label{44}
\begin{eqnarray}
&& \Gamma_{R}^{(2)}(k=0, j=0, g, 0) = \sigma^{2} \tau^{2}, \\
&& \frac{\partial\Gamma_{R}^{(2)}(k=\kappa, j=0, g, 0)}{\partial k^{2}}|_{k^{2}=\kappa^{2}} \
= 1, \\
&& \Gamma_{R}^{(4)}(k_{l}, i_{l}=0, g, 0)|_{SP} = g  , \\
&& \Gamma_{R}^{(2,1)}(k_{1}, i_{1}=0, k_{2}, i_{2}=0, Q, j'=0, g, 0)|_{\bar{SP}} = 1 ,
\end{eqnarray}
\end{subequations} 
It is important to mention that the symmetry point implies that the insertion 
momentum in last equation satisfies $Q^{2}= (k_{1}+k_{2})^{2}$. 
\par Multiplicative renormalization arguments can be most easily implemented 
when the bare theory is regularized through the ultraviolet cutoff. Indeed, 
when the normalization conditions given above are replaced into the 
renormalized vertex parts defined by 
\begin{equation}\label{45}
\Gamma_{R}^{(N,M)}(p_{n}, i_{n}, Q_{m}, i'_{m}, g, 0)= (Z_{\phi}^{(\tau)})^{\frac{N}{2}}(Z_{\phi^{2}}^{(\tau)})^{M} \Gamma^{(N,M)}(p_{n}, i_{n}, Q_{m}, i'_{m}, \lambda_{0}, 0, \Lambda),
\end{equation} 
they turn out to render them automatically finite when the regulator 
$\Lambda$ is taken to infinity. Although we mentioned the cutoff, we could 
also use another regularization scheme. In fact, we shall shift the argument 
to consider dimensionally regularized diagrams. We shall utilize 
the cutoff whenever we can successfully simplify the point under consideration.
\par Imposing that the bare theory does not depend on the momentum scale 
where the renormalized theory is defined we find a renormalization group 
equation in terms of dimensionful quantities. At the critical dimension the 
coupling constant is dimensionful just as discussed in the massive setting. 
Away from the critical dimension, similar arguments can be devised to go from 
dimensionful quantities to dimensionless amounts. 
\par Let the flow functions be defined by the 
expressions
$\beta^{(\tau)}(\kappa, g) 
= \kappa \frac{\partial g}{\partial \kappa}$, 
$\gamma_{\phi}^{(\tau)}
= \kappa \frac{\partial ln Z_{\phi}^{(\tau)}}{\partial \kappa}$ and 
$\gamma_{\phi^{2}}^{(\tau)}
= - \kappa \frac{\partial ln Z_{\phi^{2}}^{(\tau)}}{\partial \kappa}$. The 
renormalized (bare) dimensionful coupling constant is defined in terms 
of $\kappa$ as $g= \kappa^{\epsilon}u$ ($\lambda= \kappa^{\epsilon}u_{0}$), 
where $u$ ($u_{0}$) is the dimensionless renormalized coupling constant.
In order to get rid of undesirable dimensionful parameters when 
$d=4-\epsilon$, define the Gell-Mann-Low function 
$[\beta(g,\kappa)]_{GL}= -\epsilon g + \beta(g, \kappa)$. Consequently, we 
find that  
$[\beta(g,\kappa)]_{GL} \frac{\partial}{\partial g}= \beta(u) \frac{\partial}
{\partial u}$, and the renormalization group equation for the multiplicatively 
renormalized vertex parts read
\begin{eqnarray}\label{46}
&&(\kappa \frac{\partial}{\partial \kappa} + 
\beta(u) \frac{\partial}{\partial u}
- \frac{N}{2} \gamma_{\phi}^{(\tau)} + M \gamma_{\phi^{2}}^{(\tau)})
\Gamma_{R}^{(N,M)} (p_{n}, i_{n}, Q_{m}, i'_{m}, u, 0)=  0,
\end{eqnarray} 
where $\beta^{(\tau)}(u)= -\epsilon (\frac{\partial lnu_{0}^{(\tau)}}{\partial u})$, $\gamma_{\phi}^{(\tau)}(u) = \beta^{(\tau)} 
(\frac{\partial ln Z_{\phi}^{(\tau)}}{\partial u})$ 
and $\gamma_{\phi^{2}}^{(\tau)} = \beta^{(\tau)} 
(\frac{\partial ln Z_{\phi^{2}}^{(\tau)}}{\partial u})$. The combinations 
$\bar{Z}_{\phi^{2}}^{(\tau)}= Z_{\phi^{2}}^{(\tau)}Z_{\phi}^{(\tau)}$ and 
$\bar{\gamma}_{\phi^{2}}^{(\tau)} = \beta^{(\tau)} 
(\frac{\partial ln \bar{Z}_{\phi^{2}}^{(\tau)}}{\partial u})$ shall also be 
employed. We emphasize that the dynamic variable now is the 
external momentum scale where the renormalized theory is defined. The solution 
is identical to that from the ordinary $\phi^{4}$ theory and it is not going to be discussed here. As discussed in the massive theory, we employ solely the 
definitions above for the sake of determination of fixed points and other 
universal quantities via diagrammatic tools. 
\par To begin with, we write down the one-loop contribution for the four-point function, namely
\begin{eqnarray}\label{47}
I_{2}^{(\tau)} (k, i; \sigma) &=& \sigma \sum_{j=-\infty}^{\infty} 
\int d^{d-1}q \frac{1}{[(q)^{2} + (\sigma)^{2}(j+\tau)^{2}]} \nonumber
\\  
&& \times \frac{1}{[(q+k)^{2} + (\sigma)^{2}(j+ i+ \tau)^{2}]}.
\end{eqnarray}
We utilize Feynman parameters to solve the integral 
over the continous momentum using standard formulae for dimensional 
regularization. Using Eqs. (\ref{10}) and (\ref{12}) along with the identity 
$\sqrt{\pi}\Gamma(\frac{d-1}{2})S_{d-1}= \Gamma(\frac{d}{2}) S_{d}$, 
where $S_{d}$ is the area of the $d$-dimensional unit sphere, we obtain a 
result proportional to $S_{d}$. As before, we neutralize this angular 
factor appearing in each loop integral by absorbing it in a redefinition of 
the coupling constant. Collecting this steps together we determine 
the massless expression for the four-point one-loop contribution in the form
\begin{eqnarray}\label{48}
&I_{2}^{(\tau)} (k,i;\sigma) \equiv \frac{I_{2}^{(\tau)} (k,i;\sigma)}{S_{d}} = \frac{1}{\epsilon}
\bigl((1-\frac{\epsilon}{2})\int_{0}^{1} dx [x(1-x)(k^{2} 
\nonumber\\
& + \sigma^{2} i^{2})^{ - \frac{\epsilon}{2}} + \frac{\epsilon}{2} \Gamma(2-\frac{\epsilon}{2}) F^{(\tau)}_{\frac{\epsilon}{2}}(k,i;\sigma)\bigr),
\end{eqnarray} 
where 
\begin{equation}\label{49}
F^{(\tau)}_{\alpha}(k,i;\sigma)= \sigma^{-2 \alpha} \int_{0}^{1} dx 
f_{\frac{1}{2} + \alpha}\Bigl(\tau+xi, h'(k, i, \sigma) \Bigr), 
\end{equation}
and 
\begin{equation}\label{50}
h'(k, i, \sigma)= \sigma^{-1} \sqrt{x(1-x)(k^{2} + \sigma^{2} i^{2})}.
\end{equation}
Note that Eqs. (\ref{49}) and (\ref{50}) are the massless 
counterparts of the massive definitions Eqs. (\ref{17}) and (\ref{18}), 
respectively. Using the representation (\ref{13}), the above definitions 
lead to
\begin{eqnarray}\label{51}
& F^{(\tau)}_{\frac{\epsilon}{2}}(k,i;\sigma) = 4 \sigma^{-\epsilon} 
\overset{\infty}{\underset{n=1}{\sum}}\int_{0}^{1} dx cos(2\pi n(\tau + ix))
(\frac{\sigma\pi n}{[x(1-x)(k^{2}+ i^{2} \sigma^{2})]^{\frac{1}{2}}})^{\frac{\epsilon}{2}}
\nonumber \\
& \times K_{\frac{\epsilon}{2}}(2 \pi n \sigma^{-1}[x(1-x)(k^{2}+ i^{2} \sigma^{2})]^{\frac{1}{2}}).
\end{eqnarray} 
For both normalization conditions and minimal subtraction, the external 
quasi-momentum label can be taken as the zero mode value ($i=0$) without 
loss of generality, which simplifies our task. Recalling that the finite 
size correction is $O(\epsilon^{0})$ and neglecting $O(\epsilon)$ terms, 
we can rewrite Eq.(\ref{48}) as
\begin{eqnarray}\label{52}
&I_{2}^{(\tau)} (k,i=0;L)= \frac{1}{\epsilon}
\bigl(1 - \frac{\epsilon}{2} - \frac{\epsilon}{2} \int_{0}^{1} dx 
ln[x(1-x)k^{2}]\bigr) \nonumber \\
& + 2 \overset{\infty}{\underset{n=1}{\sum}} \int_{0}^{1} dx 
cos(2 \pi n \tau) K_{0}(nL[x(1-x)k^{2}]^{\frac{1}{2}}).
\end{eqnarray} 
\par We can readily take the limit $L \rightarrow \infty$ 
$(\sigma=\frac{2\pi}{L})$ by considering the 
Bessel function for large values of its argument. Its asymptotic value is 
given by $K_{\alpha}(z') = \sqrt{\frac{\pi}{2z'}} e^{-z'}[1 + O(1/z')]$ with 
$z'=nLk \sqrt{x(1-x)}\equiv nB$ and $n=1, 2,...$. Therefore, the correction 
becomes 
\begin{eqnarray}\label{53}
&& 2 \overset{\infty}{\underset{n=1}{\sum}} \int_{0}^{1} dx 
cos(2 \pi n \tau) K_{0}(nL[x(1-x)k^{2}]^{\frac{1}{2}}) = 
\sqrt{\frac{2 \pi}{Lk}} \int_{0}^{1} dx [x(1-x)]^{-\frac{1}{4}} \nonumber\\ 
&& \;\;\times\;\;\overset{\infty}{\underset{n=1}{\sum}}
cos(2 \pi n \tau) exp(-nB).
\end{eqnarray}
The remaining series is a geometric one that can be computed when we express 
the cosine in terms of the complex exponents. We can write it in terms of an 
upper bound through the inequality
\begin{equation}\label{54}
\overset{\infty}{\underset{n=1}{\sum}}
cos(2 \pi n \tau) exp(-nB)= 
\bigl[\frac{1}{1- exp(-B) \delta_{\tau,0} + exp(-B) \delta_{\tau,\frac{1}{2}}}-1\bigr] < 1.
\end{equation}
If we take directly the limit $L \rightarrow \infty$ before performing the 
integral we see that this term vanishes and multiplies the prefactor which 
is also zero. Instead, if we use the upper bound we can estimate the integral 
which in the limit $L \rightarrow \infty$ implies the 
result
\begin{eqnarray}\label{55}
&2\bigl[ \overset{\infty}{\underset{n=1}{\sum}} \int_{0}^{1} dx 
cos(2 \pi n \tau) K_{0}(nL[x(1-x)k^{2}]^{\frac{1}{2}})\bigr] <
\sqrt{\frac{2 \pi}{Lk}}[\frac{[\Gamma(\frac{3}{4})]^{2}}{\Gamma(\frac{3}{2})}] 
\nonumber\\
& \rightarrow 0.
\end{eqnarray}
Therefore, the finite size correction interpolates from the contribution for 
large but finite values of $L$ and vanishes to infinite values of $L$ even 
in the massless case. 
Sometimes, it is useful to define the bulk parametric 
integral $i(k)= \int_{0}^{1} dx ln[x(1-x)k^{2}]$, take $k=\sqrt{k^{2}}$ and 
consider the correction term at $\epsilon=0$. 
\par Let us try to comprehend the limit $L \rightarrow 0$. Using Eq.(\ref{24}) 
with the replacements $x \rightarrow z=L \sqrt{x(1-x)k^{2}}$, 
$t= \frac{2\pi \tau}{z}$  
and taking into account the above observations, we find at $i=0$ a simple 
expression useful for minimal subtraction 
\begin{eqnarray}\label{56}
&& I_{2}^{(\tau)}(k, i=0; L) = 
\frac{1}{\epsilon}
\bigl(1 - \frac{\epsilon}{2} - \frac{\epsilon}{2} i(k) \bigr) + \bigl[\gamma + \int_{0}^{1} dx ln [\frac{Lk(x(1-x))^{\frac{1}{2}}}{2\pi}]
\bigr] \nonumber\\
&&+ \frac{1}{2}\int_{0}^{1} dx \frac{1}{\sqrt{(\tau^{2} + x(1-x)[\frac{Lk}{2\pi}]^{2})}} 
+ \frac{1}{2}\sum_{n=1}^{\infty} \int_{0}^{1} dx [\frac{1}{\sqrt{((n-\tau)^{2} + x(1-x)[\frac{Lk}{2\pi}]^{2})}}- \frac{1}{n}]\nonumber\\
&& + \frac{1}{2}\sum_{n=1}^{\infty} \int_{0}^{1} dx [\frac{1}{\sqrt{((n+\tau)^{2} + x(1-x)[\frac{Lk}{2\pi}]^{2})}}- \frac{1}{n}],
\end{eqnarray} 
whereas computing it at the symmetry point convenient for normalization 
conditions ($k^{2}=1$), we obtain
\begin{eqnarray}\label{57}
&& I_{2SP}^{(\tau)}(\sigma) = 
\frac{1}{\epsilon}
\bigl(1 +\frac{\epsilon}{2}\bigr) + \bigl[\gamma + \int_{0}^{1} dx ln [\frac{L(x(1-x))^{\frac{1}{2}}}{2\pi}]
\bigr] \nonumber\\
&&+ \frac{1}{2}\int_{0}^{1} dx \frac{1}{\sqrt{(\tau^{2} + x(1-x)[\frac{L}{2\pi}]^{2})}} 
+ \frac{1}{2}\sum_{n=1}^{\infty} \int_{0}^{1} dx [\frac{1}{\sqrt{((n-\tau)^{2} + x(1-x)[\frac{L}{2\pi}]^{2})}}- \frac{1}{n}]\nonumber\\
&& + \frac{1}{2}\sum_{n=1}^{\infty} \int_{0}^{1} dx [\frac{1}{\sqrt{((n+\tau)^{2} + x(1-x)[\frac{L}{2\pi}]^{2})}}- \frac{1}{n}].
\end{eqnarray} 
\par Except for the integrals, the $L$-dependence in last equation has 
pretty much the same form as Eq.(\ref{25}) has in $r^{-1}= \frac{L}{2 \pi \xi}$ described in the massive case. In that case the theory renormalized at 
``mass'' $\xi=1$ (fixed and finite bulk correlation length) and zero external 
momenta is completely analogous to our renormalized massless theory at the 
symmetry $\kappa^{2}=1$ (infinite bulk correlation 
length). After those choices we just have to recall the variable $L$ is 
dimensionless. Except for minor modifications like 
the extra integral on the Feynman parameter $x$ the discussion of the various 
terms parallels that for the massive case. 
\par In order to study the limit 
$L \rightarrow 0$, let us start with the last two integrals. It is licit to 
take $L=0$ inside both of them, such that their summation produces precisely 
Eq.(\ref{27}). The first integral is 
straightforward and its result is $ln\bigl[\frac{L}{4 \pi}\bigr] -1$. We are 
then left with the task of evaluating the second integral. The 
identity \cite{GR} 
\begin{equation}\label{58}
\int \frac{dx}{\sqrt{a + bx + cx^{2}}}= \frac{-1}{\sqrt{-c}} 
arcsin\bigl[\frac{2cx+b}{\sqrt{-\Delta}}\bigr],
\end{equation}
is valid for $c<0$ and $\Delta<0$, where $\Delta= 4ac - b^{2}$. Performing 
the identifications $a=\tau^{2}$ and $b=-c= \frac{L^{2}}{4\pi^{2}}$, we get 
to
\begin{equation}\label{59}
\int_{0}^{1} \frac{dx}{\sqrt{\tau^{2} + \frac{L^{2}}{4 \pi^{2}}x(1-x)}}=
\frac{4\pi}{L} 
arcsin\bigl[\frac{1}{\sqrt{1+ \frac{16 \pi^{2}\tau^{2}}{L^{2}}}}\bigr].
\end{equation} 
Collecting these steps together, we can rewrite the one-loop contribution to 
the four-point function in the form
\begin{eqnarray}\label{60}
&& I_{2SP}^{(\tau)}(\sigma) = 
\frac{1}{\epsilon}
\bigl(1 +\frac{\epsilon}{2}\bigr) + \bigl[\gamma 
+ (2ln2-1)\delta_{\tau,\frac{1}{2}} + ln \bigl[\frac{L}{4 \pi}\bigr] -1 \bigr] 
\nonumber\\
&& + \frac{2\pi}{L} 
arcsin\bigl[\frac{1}{\sqrt{1+ \frac{16 \pi^{2}\tau^{2}}{L^{2}}}}\bigr].
\end{eqnarray}
\par Last equation shows that last term inside the finite size contribution 
for periodic boundary condition $\tau=0$ becomes exactly $\frac{\pi^{2}}{L}$ 
which differs from the expression in the massive case (with fixed $\xi=1$) 
by a factor of $\pi$ due to the effect of performing the integral over the 
Feynman parameter $x$. The new situation occurs for antiperiodic boundary 
condition ($\tau=\frac{1}{2}$) in the last term, whose limit $L \rightarrow 0$ 
becomes $\frac{\pi}{L^{2}}$. The logarithm is still there just as before in 
the massive case, but now these two different power-law behavior present in 
both boundary conditions are dominant in the $L \rightarrow 0$ limit. 
Therefore, the finite size correction is generically enhanced for both 
boundary conditions in the critical massless theory consistent with the 
enhancement of fluctuations at this regime. At the 
critical point crossover starts earlier in antiperiodic than in periodic 
boundary condition as evidenced by our analytical result in above equation. 
Thus, the massless case has nontrivial aspects in comparison with the massive 
case, as shown here for the first time, since the second type of crossover 
discussed in Sec. II for $t>0$ in antiperiodic boundary condition is now 
absent for $t=0$. 
\par Provided we stay away from the crossover regions characterized by very 
small values of $L$, the two descriptions are almost equivalent, even though 
the equivalence is not complete, as far as the finite size contribution is 
concerned. The $\epsilon$-expansion is well defined in both situations, if the 
crossover regions are precluded from our analysis. What is really remarkable 
from the massless and massive analysis is the bulk correlation length 
independence of the finite size correction. From last equation, finite values 
of $L$ persist in the correction even when the starting point of the massless 
theory corresponds to $\frac{L}{\xi}\rightarrow 0$. Thus, region c) is 
available to our scrutiny. Consequently, the previous 
phenomenological conjecture that the massless limit cannot be understood in 
terms of $\epsilon$-expansion is unfounded. From now on, we 
are going to consider finite (but not too small values) of $L$, since the 
finite size correction has a good behavior and $\epsilon$-expansion methods 
can be utilized without further problems. 
\par We leave the task of computing the higher loop integrals to Appendix B. 
Here we simply quote the results. The massless counterpart of the integral 
which contributes to the four-point function at two-loops can be extracted 
from Eq.(\ref{29}) by setting $\mu=0$, namely 
\begin{eqnarray}\label{61}
&&I_{4}(k_{1},k_{2},k_{3}, k_{4},i_{1},i_{2},i_{3}, i_{4},\sigma) = \sigma^{2}\sum_{j_{1},j_{2}=-\infty}^{\infty}
\int \frac{d^{d-1}{q_{1}}d^{d-1}q_{2}}
{\left(q_{1}^{2} + \sigma^{2}(j_{1} + \tau)^2 \right)} \nonumber\\
&& \;\times \frac{1}{\left(q_{2}^{2} + \sigma^{2}(j_{2} + \tau)^2 \right)[(q_{1} - q_{2} + k_{3})^{2} + \sigma^{2}(j_{1} - j_{2} + i_{3} + \tau)^2]} \nonumber\\
&&\;\;\times \frac{1}{[(P - q_{1})^{2} + \sigma^{2}(p - j_{1} 
+ \tau)^2]}\;\;,
\end{eqnarray}
\par At zero external quasimomenta, there is no loss of generality to 
approaching the minimal subtraction scheme with arbitrary external momenta or 
normalization conditions with fixed nonvanishing external momenta. From our 
discussion in Appendix B we obtain 
\begin{equation}\label{62}
I_{4}^{(\tau)} (k_{i},0;\sigma)= \frac{1}{2 \epsilon^{2}}
\bigl(1-\frac{\epsilon}{2}- \epsilon i(P) + \epsilon F_{\epsilon}^{(\tau)}
(\frac{PL}{2\pi}, 0)\bigr).
\end{equation}
At the symmetry point $P^{2}=\kappa^{2}=1$ it takes the simpler form 
by setting $\epsilon=0$ into the finite size correction
\begin{equation}\label{63}
I_{4SP}^{(\tau)} (\sigma)= \frac{1}{2 \epsilon^{2}}
\bigl(1+\frac{3\epsilon}{2} + \epsilon F_{0}^{(\tau)}
(\sigma)\bigr).
\end{equation}
\par Analogously, the integrals contributing to the two-point function at 
two- and three-loops can be read off from Eqs. (\ref{31}) and (\ref{32}) 
at $\mu=0$, respectively, see Appendix B. $I_{3}(k, i=0;\sigma)$ in a form 
appropriate to minimal subtraction reads
\begin{equation}\label{64}
I_{3}^{(\tau)} (k,\sigma)= -\frac{1}{8\epsilon}
\bigl((k^{2} + \sigma^{2} \tau^{2})[1+\frac{\epsilon}{4} - 2\epsilon i_{3}(k^{2} + \sigma^{2} \tau^{2})]
- 2 \epsilon \tilde{F}_{\epsilon}^{(\tau)}(k,i=0;\sigma)- 4 \epsilon 
F_{\frac{\epsilon}{2},1}(k, i=0; \sigma)\bigr). 
\end{equation}
In normalization conditions the derivative of this integral with respect 
to $k^{2}$ computed at the symmetry point $k^{2}=1$ can be written as  
\begin{equation}\label{65}
I_{3SP}^{\prime (\tau)} (\sigma)= -\frac{1}{8\epsilon}
\bigl(1+\frac{5\epsilon}{4} - 2\epsilon W_{0}(\sigma)\bigr). 
\end{equation}
\par Using a similar reasoning, the solution for $I_{5}$ appropriate within 
minimal subtraction of dimensional poles was found to be
\begin{equation}\label{66}
I_{5}^{(\tau)} (k,\sigma)= -\frac{1}{6\epsilon^{2}}
\bigl((k^{2} + \sigma^{2} \tau^{2})[1+\frac{\epsilon}{2} - 3\epsilon i_{3}(k^{2} + \sigma^{2} \tau^{2})]
- 3 \epsilon \check{F}_{\frac{3\epsilon}{2}}^{(\tau)}(k,i=0;\sigma) 
- 6 \epsilon F_{\frac{\epsilon}{2},1}(k, i=0; \sigma)\bigr). 
\end{equation}
On the other hand, its derivative in relation to $k^{2}$ at the symmetry 
point $k^{2}=1$ is required when applying normalization conditions as our 
renormalization scheme. Hence,  
\begin{equation}\label{67}
I_{5SP}^{\prime (\tau)} (\sigma)= -\frac{1}{6\epsilon^{2}}
\Bigl(1+ 2\epsilon - 3\epsilon W_{0}(\sigma)\Bigr). 
\end{equation}
Note that we can safely replace the values of the subscript of the 
additional functions appearing as finite size corrections at $\epsilon=0$ 
into Eqs. (\ref{62}), (\ref{64}) and (\ref{66}) when we employ the minimal 
subtraction scheme. This will facilitate the computations of the normalization 
functions, since many different functions become identical at $\epsilon=0$. 
We have now all the required integrals to compute critical exponents in 
normalization conditions or minimal subtraction, to which we turn our 
attention next.

\section{Critical exponents from the massless approach}
\par The results displayed in last section can be substantiated in the 
calculation of the critical exponents $\nu$ and $\eta$ at two- and 
three-loop order, respectively, through 
diagrammatic (perturbative) methods. We are going to compute the critical 
indices and show that they are independent of the renormalization scheme, 
using either normalization conditions or minimal subtraction. 

\subsection{Normalization conditions}
\par As before, we write the dimensionless bare coupling constant and 
normalization function $Z_{\phi}$, $\bar{Z}_{\phi^{2}}$ as power 
series in the dimensionless renormalized coupling constant as 
$u_{0}^{(\tau)} = u (1 + a_{1}^{(\tau)}u + a_{2}^{(\tau)}u^{2})$, 
$Z_{\phi}^{(\tau)} = 1 + b_{2}^{(\tau)} u^{2} + b_{3}^{(\tau)} u^{3}$ and 
$\bar{Z}_{\phi^{2}}^{(\tau)} = 1 + c_{1}^{(\tau)} u + c_{2}^{(\tau)} u^{2}$.
A considerable labor can be saved by noting that the structure of these 
equations are the same as those previously discussed in Sec.III for the 
massive case. Indeed, using Eqs.(\ref{35}) from Sec.III and replacing the 
values of the massless integrals computed at the symmetry point, we can 
determine the above coefficients in the massless theory as well. Following 
this trend, we encounter the following values for the coefficients
\begin{subequations}\label{68}
\begin{eqnarray}
&& a_{1}^{(\tau)} = \frac{(N+8)}{6 \epsilon}[1 + \frac{\epsilon}{2}
+ \frac{\epsilon}{2} F_{0}^{(\tau)}(\sigma)] ,\label{68a}\\
&& a_{2}^{(\tau)} = (\frac{N+8}{6 \epsilon})^{2}
+ \frac{2N^{2}+23N+86}{72 \epsilon} 
+ \bigl(\frac{N+8}{6}\bigr)^{2} \frac{F_{0}^{(\tau)}(\sigma)}{\epsilon},
\label{68b}\\
&& b_{2}^{(\tau)} = -\frac{(N+2)}{144 \epsilon}[1 + \frac{5\epsilon}{4} 
- 2 \epsilon W_{0}^{(\tau)}(\sigma)], \label{68c}\\
&& b_{3}^{(\tau)} = -\frac{(N+2)(N+8)}{1296 \epsilon^{2}}[1 
+ \frac{5\epsilon}{4} + \frac{3\epsilon}{2}F_{0}^{(\tau)}(\sigma)], 
\label{68d}\\
&& c_{1}^{(\tau)} = \frac{(N+2)}{6 \epsilon}[1 +\frac{\epsilon}{2} 
+ \frac{\epsilon}{2} F_{0}^{(\tau)}(\sigma)],\label{68e} \\
&& c_{2}^{(\tau)} = \frac{(N+2)(N+5)}{36 \epsilon^{2}}
+ \frac{2N^{2}+11N+14}{72 \epsilon} 
+ \frac{(N^{2} + 7N +10)}{36 \epsilon} F_{0}^{(\tau)}(\sigma) .\label{68f}
\end{eqnarray}
\end{subequations} 
\par Now, introducing the coefficients $a_{1}^{(\tau)}$ 
as well as $a_{2}^{(\tau)}$ appropriate to the massless 
formulation into Eqs.(\ref{36}) from Sec. III, we first determine the flow 
function in the zero mass limit which yields
\begin{equation}\label{69}
\beta^{(\tau)}(u) = -\epsilon u 
+ \bigl(\frac{N+8}{6}\bigr)(1 + \frac{\epsilon}{2}
+ \frac{\epsilon}{2} F_{0}^{(\tau)}(\sigma))u^{2} - \frac{3N+14}{12} u^{3}.
\end{equation} 
The attractive nontrivial infrared fixed point at two-loop order is found 
out from the condition $\beta^{(\tau)}(u^{*})=0$, namely
\begin{equation}\label{70}
u^{*}= \bigl(\frac{6}{N+8}\bigr) \epsilon 
\bigl[1+\bigl(\frac{9N+42}{(N+8)^{2}} - \frac{1}{2} - \frac{1}{2} F_{0}^{(\tau)}(\sigma)\bigr)\epsilon \bigr].
\end{equation} 
\par Let us now utilize Eqs. (\ref{39a}) and (\ref{39b}) in order to 
determine the Wilson functions for the massless formalism. Inserting the 
coefficients determined above into that equation and proceeding along the same 
lines as before, we get the expressions
\begin{subequations}\label{71}
\begin{eqnarray}
&& \gamma_{\phi}^{(\tau)} = u^{2} \frac{(N+2)}{72}[1 + \frac{5\epsilon}{4} 
- 2 \epsilon W_{0}^{(\tau)}(\sigma) - \frac{(N+8)}{12}(1 - 
4W_{0}^{(\tau)}(\sigma) 
- 2 F_{0}^{(\tau)}(\sigma))u],\label{71a}\\
&& \bar{\gamma}_{\phi^{2}}^{(\tau)} = \frac{(N+2)}{6}(1 + \frac{\epsilon}{2} 
+ \frac{\epsilon}{2} F_{0}^{(\tau)}(\sigma))u - \frac{(N+2)}{12} u^{2}.
\label{71b}
\end{eqnarray}
\end{subequations}
Computing these functions at the nontrivial infrared fixed point $u^{*}$, 
we find that, i) $\eta=\gamma_{\phi}^{(\tau)}(u^{*})$ is identical to 
Eq.(\ref{41}) and ii) $\bar{\gamma}_{\phi^{2}}^{(\tau)} (u^{*})$ is equal to 
the expression in Eq.(\ref{42}), which leads to the exponent $\nu$ from Eq.(\ref{43}). 
\par The massive treatment in the ultraviolet regime is therefore completely 
equivalent to the massless framework at the infrared region, for they 
originate the same critical indices, even though the intermediate steps 
are completely distincts in the two formalisms, in compliance with 
universality. This thorough treatment of massless fields shall be concluded 
in a moment with the computation using minimal subtraction.      

\subsection{Minimal subtraction}
\par Here we are not going to calculate explicitly the critical
exponents. Instead, we are going to calculate the fixed point as well as
the functions $\gamma_{\phi}^{(\tau)}$ and  $\bar{\gamma}_{\phi^{2}}^{(\tau)}$
at the fixed point. As these functions at the fixed point are
universal, they should be equal to the ones obtained using
normalization conditions, leading to the same exponents in either
renormalization scheme.
\par The dimensionless bare couplings and the renormalization functions
are defined in minimal subtraction by
\begin{subequations}\label{72}
\begin{eqnarray}
&& u_{0}^{(\tau)} = u[1 + \sum_{i=1}^{\infty} a_{i}^{(\tau)}(\epsilon)
u^{i}], \label{72a}\\
&& Z_{\phi}^{(\tau)} = 1 + \sum_{i=1}^{\infty} b_{i}^{(\tau)}(\epsilon)
u^{i}, \label{72b}\\
&& \bar{Z}_{\phi^{2}}^{(\tau)} = 1 + \sum_{i=1}^{\infty} c_{i}^{(\tau)}
(\epsilon)u^{i}.\label{72c}
\end{eqnarray}
\end{subequations}
The renormalized vertices
\begin{subequations}\label{73}
\begin{eqnarray}
&& \Gamma_{R}^{(2)}(k, u, \kappa) = Z_{\phi}^{(\tau)}
\Gamma^{(2)}(k, u_{0}^{(\tau)},\kappa ), \label{73a}\\
&& \Gamma_{R}^{(4)}(k_{i}, u, \kappa) = \bigl(Z_{\phi}^{(\tau)}\bigr)^{2} \Gamma^{(4)}(k_{i}, u_{0}^{(\tau)},\kappa), \label{73b}\\
&& \Gamma_{R}^{(2,1)}(k_{1}, k_{2}, p; u, \kappa) = \bar{Z}_{\phi^{2}}^{(\tau)}
\Gamma^{(2,1)}(k_{1}, k_{2}, p, u_{0}^{(\tau},\kappa),\label{73c}
\end{eqnarray}
\end{subequations}
should be finite when $\epsilon\rightarrow 0$ to any desired order in
$u$. Observe that the external momenta into the bare vertices
are mutiplied by $\kappa^{-1}$ and all the external quasimomenta of the 
diagrams are set to zero in order to simplify matters. The coefficients 
$a_{i}^{(\tau)}(\epsilon)$, 
$b_{i}^{(\tau)}(\epsilon)$ and $c_{i}^{(\tau)}(\epsilon)$ are obtained
by requiring that the poles in $\epsilon$ be minimally subtracted.
The bare vertices can now be expressed as
\begin{subequations}\label{74}
\begin{eqnarray}
&& \Gamma^{(2)}(k, u_{0}^{(\tau)}, \kappa) =
k^{2}(1- B_{2}^{(\tau)} (u_{0}^{(\tau)})^{2} + B_{3}^{(\tau)}(u_{0}^{(\tau)})^{3}), \label{74a}\\
&& \Gamma^{(4)}(k_{i}, u_{0}^{(\tau)}, \kappa) =
\kappa^{\epsilon} u_{0}^{(\tau)}
[1- A_{2}^{(\tau)} u_{0}^{(\tau)}
+ (A_{2}^{(\tau)\;(1)} + A_{2}^{(\tau)\;(2)})(u_{0}^{(\tau)})^{2}],\label{74b}\\
&& \Gamma^{(2,1)}(k_{1}, k_{2}, p;
u_{0}^{(\tau)}, \kappa) = 1 - C_{1}^{(\tau)} u_{0}^{(\tau)}
+ (C_{2}^{(\tau)\;(1)} + C_{2}^{(\tau)\;(2)}) (u_{0}^{(\tau)})^{2}.\label{74c}
\end{eqnarray}
\end{subequations}
We recognize that $B_{2}^{(\tau)}$ is proportional to the integral 
$I_{3}^{(\tau)}$ and
$B_{3}^{(\tau)}$ is proportional to $I_{5}^{(\tau)}$. In the remainder 
we shall suppress the upper indices in the integrals referring to the 
boundary condition, but keeping them implicitly. Explicitly, the 
coefficients can be written in terms of integrals like
\begin{subequations}\label{75}
\begin{eqnarray}
&& A_{1}^{(\tau)} = \frac{(N+8)}{18}[ I_{2}(\frac{k_{1} + k_{2}}{\kappa}) 
+  I_{2}(\frac{k_{1} + k_{3}}{\kappa}) + I_{2}(\frac{k_{2} + k_{3}}
{\kappa})] ,\label{75a}\\
&& A_{2}^{(\tau)\;(1)} = \frac{(N^{2} + 6N + 20)}{108}
[I_{2}^{2}(\frac{k_{1} + k_{2}}{\kappa})
+  I_{2}^{2}(\frac{k_{1} + k_{3}}
{\kappa}) + I_{2}^{2}(\frac{k_{2} + k_{3}}
{\kappa})] ,\label{75b}\\
&& A_{2}^{(\tau)\;(2)} = \frac{(5N + 22)}{54}
[I_{4}( \frac{k_{i}}{\kappa}) + 5   \;permutations] ,\label{75c}\\
&& B_{2}^{(\tau)} = \frac{(N+2)}{18}I_{3}(\frac{k}{\kappa}) ,\label{75d}\\
&& B_{3}^{(\tau)} =
\frac{(N+2)(N+8)}{108}I_{5}(\frac{k}{\kappa}) ,\label{75e}\\
&& C_{1}^{(\tau)} = \frac{N+2}{18}[ I_{2}(\frac{k_{1} + k_{2}}
{\kappa}) +  I_{2}(\frac{k_{1} + k_{3}}
{\kappa}) + I_{2}(\frac{k_{2} + k_{3}}
{\kappa})] ,\label{75f}\\
&& C_{2}^{(\tau)\;(1)} = \frac{(N+2)^{2}}{108}
[I_{2}^{2}(\frac{k_{1} + k_{2}}{\kappa})
+  I_{2}^{2}(\frac{k_{1} + k_{3}}
{\kappa}) + I_{2}^{2}(\frac{k_{2} + k_{3}}
{\kappa})] ,\label{75g}\\
&& C_{2}^{(\tau)\;(2)} = \frac{N+2}{36}[I_{4}(\frac{k_{i}}{\kappa}) + 5   \;permutations].\label{75h}
\end{eqnarray}
\end{subequations}
\par Firstly substitute Eqs. (\ref{75}) inside Eqs.(\ref{74}). Next, 
utilize Eq.(\ref{72a}) into Eqs.(\ref{74}). Finally, impose that the 
renormalized vertex parts expressed as Eqs.(\ref{73}) are finite via 
minimal subtraction of dimensional poles. Interestingly, all the logarithmic
integrals in the external momenta as well as the finite size corrections 
appearing in $I_{2}, I_{3}, I_{4}$, and $I_{5}$ cancell each other in the 
algorithm of renormalization. This results in the determination of the 
coefficients in minimal subtraction, or in other words 
\begin{subequations}\label{76}
\begin{eqnarray}
u_{0}^{(\tau)} &=& u(1 + \frac{(N+8)}{6 \epsilon} u
+ [\frac{(N+8)^{2}}{36 \epsilon^{2}} - \frac{(3N+14)}{24
\epsilon}] u^{2}), \label{76a}\\
Z_{\phi}^{(\tau)} &=& 1 - \frac{N+2}{144 \epsilon} u^{2}
+ [-\frac{(N+2)(N+8)}{1296  \epsilon^{2}} + \frac{(N+2)(N+8)}{5184
\epsilon}] u^{3}, \label{76b}\\
\bar{Z}_{\phi^{2}}^{(\tau)} &=& 1 + \frac{N+2}{6\epsilon} u
+ [\frac{(N+2)(N+5)}{36\epsilon^{2}} - \frac{(N+2)}{24\epsilon}] u^{2}).
\label{76c}
\end{eqnarray}
\end{subequations}
What is amazing is that in minimal subtraction the above mentioned 
renormalization functions 
{\it do not depend on the boundary condition explicitly}, since all that 
dependence cancelled out naturally, i.e., they do not appear in the right 
hand side of the above equations.  
\par Furthermore, from the renormalization functions one can obtain:
\begin{eqnarray}
\gamma_{\phi}^{(\tau)} &=& \frac{(N+2)}{72}u^{2}
- \frac{(N+2)(N+8)}{1728}u^{3}, \label{77}\\
\bar{\gamma}_{\phi^{2}}^{(\tau)} &=& \frac{(N+2)}{6} u
[ 1 - \frac{1}{2} u]. \label{78}
\end{eqnarray}
\par The fixed point is defined by $\beta^{(\tau)}(u^{*}) =
0$. Then,  we find:
\begin{equation}\label{79}
u^{\ast}=\frac{6}{8 + N}\,\epsilon\Biggl\{1 + \epsilon
\,\Biggl[\frac{(9N + 42)}{(8 + N)^{2}}\Biggr]\Biggr\}\;\;.
\end{equation}
\par Substitution of this result into the renormalization constants
will give at the fixed point $\gamma_{\phi}^{\ast\;(\tau)}= \eta$, 
whereas, in addition, we have
\begin{equation}\label{80}
\bar{\gamma^{\ast}}_{\phi^{2}}^{(\tau)} = \frac{(N+2)}{(N+8)} \epsilon
[ 1 + \frac{6(N+3)}{(N+8)^{2}} \epsilon].
\end{equation}
This expression is equal to Eq.(\ref{42}) obtained using the massive method 
and consequently lead to the same exponent $\nu$ from Eq.(\ref{43}). Notice 
that Eqs.(\ref{76})-(\ref{80}) are the same as their counterpart obtained in 
minimal subtraction for the usual bulk theory. All the 
construction of renormalization schemes developed in the present work are 
thus consistent with universality, which states that critical exponents 
(among other universal quantities) are scheme independent as we have shown 
herein. 
 
\section{Discussion of the results and conclusion}
\par We have computed critical exponents at higher loop order from finite size 
layered systems subject to periodic and antiperiodic boundary conditions on the limiting surfaces of the slab (parallel plate) geometry by defining a 
one-particle irreducible ($1PI$) vertex parts formalism to the previous 
field-theoretic framework of Green functions for those systems introduced by 
Nemirovsky and Freed earlier. In order to do that, we determine normalization 
functions as well as fixed points and show that they depend on the boundary 
conditions whenever we use normalization conditions either in the massive or 
massless methods. In minimal subtraction, however, we find that these 
quantities are independent of the boundary condition.
\par We confirm that for large values of $L$ and in the 
$L \rightarrow \infty$ all the finite size corrections are under control and 
the critical exponents obtained in this way are identical to those from 
$d$-dimensional (bulk) universality class. In the case of periodic boundary 
conditions we proved that dimensional 
crossover only occurs at very small values of $L$, the behavior of 
the finite size correction is proportional to $L^{-1}$ in the massive and 
massless cases, being independent of the value of the bulk correlation length, 
although the coefficient of thies term is larger in the massless case. This 
extends the previous analysis by $NF$ performed solely in the massive case. 
As far as the crossover regimes are concerned, antiperiodic boundary 
conditions  do not present the simple behavior from $PBC$. In addition to 
the ``dimensional crossover'' previously discussed by $NF$ below the critical 
temperature $t<0$, we have found a new regime of crossover for $ABC$ which 
exists only for $t>0$ characterized by much shorter values of $L$ than its 
dimensional crossover counterpart occurring in $PBC$, i.e., when the lattice 
constant is of the same order of magnitude of the distance between the 
limiting plates. Moreover, the crossover in the massless case $t=0$ for 
$ABC$ starts earlier than in the massive case. Actually, the finite size 
correction is proportional to $L^{-2}$ in that case, which starts to modify 
the bulk critical behavior for larger values of $L$ than its counterpart in 
the $PBC$ case. Thus, fluctuations at the critical point enhance the effect 
of crossover in finite systems in these simple boundary conditions.

\par Let us discuss the connection between one previous 
two-loop calculation using $NF$ formulation with our work of higher loop 
integrals computed in whole detail in Appendix B. Actually, Krech and Dietrich 
Ref.\cite{KD1} used $NF$ formulation for massless fields in their computation 
of Casimir amplitudes (see Appendix A therein). However, only one two loop 
diagram was computed for the free energy, which actually corresponds to a 
squared tadpole diagram. But this is equivalent to a one loop computation, 
since normalization constants, fixed points, etc., were computed only at 
one-loop order. Unfortunately, perhaps the lack of a better representation 
for the function $f_{\alpha}(a,b)$ at the time of the writing of that paper 
prevented them to obtain simple answers in terms of elementary primitive 
functions. Consequently, they abandoned $NF$ method in momentum space in 
their subsequent work with their collaborators. The work presented here, on 
the other hand, permits us to go beyond the simple conclusions of previous 
analysis: even though the exponents are identical to those from the bulk 
system when we avoid the crossover region which is certainly not too exciting, 
the crossover regimes assessed by the analytical results described in the 
present paper sheds new light on the fundamental difference between massive 
and massless regimes to finite size systems criticality, and how fluctuations 
enhance the effect of finiteness in the latter. Moreover, a consistent 
description in terms of massless fields with inequivalent crossover regions 
in comparison with the massive case for $ABC$ is 
certainly a step forward which cannot be underestimated. Since the 
consistency of the critical regime for both boundary conditions implies that 
phenomenological scaling relying on the failure of $\epsilon$-expansion results in the region $\frac{L}{\xi} \rightarrow 0$ is incorrect, the present study 
should be considered the starting point to widen the subject and put it on 
new grounds, such as computing higher order universal quantities like 
amplitudes in order to improve our present knowledge of critical finite 
systems.        
\par Away from the crossover regimes, we have shown a complete equivalence 
between the formulations using either massive or massless fields, where the 
renormalized mass scale plays the analogue role of the external momenta scale 
used to fix the (new) massless theory in normalization conditions. A step 
further is the minimal subtraction treatment for massless fields.
\par Thus, our critical exponents results confirm the previous expectations 
pointed out by Nemirovsky and Freed concerning a behavior identical to those 
describing the bulk system in higher order loop computations. Contrary to 
previous conjectures, it is 
{\it not} the boundedness variable $\frac{L}{\xi}<1$ which makes the 
$\epsilon$-expansion invalid, but small values of $L$ decreasing below a 
given threshold which are responsible for crossover. This crossover 
description is far from being completely figured out, but the resources 
developed in our trend here should be encouraging to tackle this problem. 
We hope our discussion in the present work can serve as an introduction to 
this subject and might be valuable when new perturbative methods to treat 
the limit $L \rightarrow 0$ become available.
\par A rather interesting topic is to extend the $1PI$ method at two-loop 
order to treat systems with Neumann and Dirichlet boundary conditions, since 
they are more appealing from the phenomenological viewpoint. They characterize 
free surfaces \cite{D}, which disturb further the system due to the breaking 
of translational invariance along the finite directions. First, the 
quasi-momentum are not conserved in the treatment of these boundary 
conditions. Consequently, in order to renormalize the theory we have to 
introduce distinct external fields, one in the bulk and other in the limiting 
surfaces. This implies that the surface parameter becomes relevant and 
requires a new normalization function to renormalize it. In spite of these 
additional aspects, we expect that this topic can be investigated along a 
similar line of reasoning to that employed in the present work. 
\par Another intriguing perspective is to consider the finite size approach 
to competing systems of the Lifshitz type \cite{L1,L2,CL}. 
It remains to be seen if the competing axes with 
arbitrary momentum powers permit exact results when the finite size 
direction points along any of them. The last few years have witnessed 
promising new applications of this kind of field theory from quantum field 
theory to quantum gravity and cosmology. The most direct 
application is to study aspects of space(time)s with one compact spatial 
dimension, called ``Lifshitz space(time)'', e. g., in the Horava-Lifshitz 
theory of gravity \cite{Horava} and other simpler quantum field theories 
\cite{Anselmi}. 
\par Finally, it is possible that the results obtained in the present paper 
must be used to update certain computations of amplitude ratios of certain 
thermodynamical potentials. Other aspects like extension of the present 
method in the analysis of semi-infinite systems are also worthwhile.
\section{Acknowledgments}
JBSJ acknowledges financial support by CNPq from Brazil.

\appendix
\section{Higher order massive integrals in dimensional 
regularization}
\par Since the relevant one-loop integral integral contributing to the 
four-point was discussed in detail in the main text, we shall discuss only 
two- and three-loop diagrams, using the one-loop result extracted from the 
text whenever possible.  
\par The required integrals are computed at unity mass, which makes 
$r\equiv\frac{\sigma}{\mu}=\frac{2\pi}{L}$ . The simpler 
contributions come from the two- and three-loop diagrams of the two-point 
function, respectively, given by the following expressions
\begin{eqnarray}\label{A1}
&& I_{3}(k,i,\sigma) = \sigma^{2}\sum_{j_{1},j_{2}=-\infty}^{\infty} 
\int \frac{d^{d-1}{q_{1}}d^{d-1}q_{2}}
{\left( q_{1}^{2} + \sigma^{2}(j_{1}+\tau)^2 + 1 \right)
\left( q_{2}^{2} + \sigma^{2}(j_{2}+\tau)^2 + 1 \right)} \nonumber\\
&& \times \;\;\; \frac{1}{[(q_{1} + q_{2} + k)^{2} 
+ \sigma^{2}(j_{1} + j_{2}+ i + \tau)^{2} + 1]}\;\;,
\end{eqnarray}
\begin{eqnarray}\label{A2}
&& I_{5}(k,i,\sigma) = \sigma^{2}\sum_{j_{1},j_{2}=-\infty}^{\infty}
\int \frac{d^{d-1}{q_{1}}d^{d-1}q_{2}d^{d-1}q_{3}}
{\left(q_{1}^{2} + \sigma^{2}(j_{1}+\tau)^{2} + 1 \right) 
\left( q_{2}^{2} +  \sigma^{2}(j_{2}+\tau)^{2} 
+ 1 \right)} \nonumber\\
&& \times \frac{1}{\left( q_{3}^{2} +  \sigma^{2}(j_{3}+\tau)^{2} 
+ 1 \right)[ (q_{1} + q_{2} + k)^{2} 
+ \sigma^{2}(j_{1}+j_{2}+i+\tau)^{2} + 1]} \nonumber\\
&& \times \frac{1}{[(q_{1} + q_{3} + k)^{2} 
+ \sigma^{2}(j_{1}+j_{3}+i+\tau)^{2} + 1]}.
\end{eqnarray}
\par The remaining integral is the nontrivial contribution to the four-point 
function at two loops, namely 
\begin{eqnarray}\label{A3}
&&I_{4}(k_{1},k_{2},k_{3}, k_{4},i_{1},i_{2},i_{3}, i_{4},\sigma) = \sigma^{2}\sum_{j_{1},j_{2}=-\infty}^{\infty}
\int \frac{d^{d-1}{q_{1}}d^{d-1}q_{2}}
{\left(q_{1}^{2} + \sigma^{2}(j_{1} + \tau)^2 + 1 \right)} \nonumber\\
&& \;\times \frac{1}{\left(q_{2}^{2} + \sigma^{2}(j_{2} + \tau)^2 + 1\right)[(q_{1} - q_{2} + k_{3})^{2} + \sigma^{2}(j_{1} - j_{2} + i_{3} + \tau)^2
+ 1]} \nonumber\\
&&\;\;\times \frac{1}{[(P - q_{1})^{2} + \sigma^{2}(p - j_{1} 
+ \tau)^2 + 1]}\;\;,
\end{eqnarray}
where $P=k_{1}+k_{2}$ is the external momenta along the plates and 
$p=i_{1}+i_{2}$ is a discrete ``external'' quasi-momentum label. For 
convenience, we shall compute all the integrals with all external 
quasi-momenta set to zero.
\par The systematics to solve $I_{3}$ and $I_{5}$ is very similar: in the 
former there appears a four-point one-loop subdiagram, whilst in the latter a 
squared of that object takes place. We first solve the internal bubble(s) 
belonging to $I_{3}$ ($I_{5}$), and use Feynman parameters to solve the 
external resulting bubble. It is important to divide each loop integral 
by the unit area of the $d$-dimensional sphere $S_{d}$, a standard 
procedure within this technique.
\par The objects required to our purposes are the derivative of those 
integrals with respect to the external momenta, setting the external 
momenta at the symmetry point in the end of the process. 
\par Let us apply this general strategy to perform a detailed computation of 
$I^{'}_{3}$ at zero external momentum and quasi-momentum. First, set the 
external quasi-momentum $i=0$ inside Eq.(\ref{A1}) in order 
to simplify our task. Let us rewrite this integral 
in terms of the one-loop subdiagram $I_{2}^{(\tau)}(k,i=0,r)$ in the form
\begin{eqnarray}\label{A4}
&& I^{(\tau)}_{3}(k,i=0;\sigma) = \sigma \sum_{j=-\infty}^{\infty} 
\int d^{d-1}q \frac{I_{2}^{(\tau)}(q+k,j;\sigma)}{[(q)^{2} + \sigma^{2}(j+\tau)^{2}
+1]}. 
\end{eqnarray} 
Sometimes, it is appropriate to perform the change of variables $q'=q+k$, 
such that the external momenta is exchanged to the external subdiagram; 
see below. Now using Eq.(\ref{16}) along with the definitions Eqs.(\ref{17}), 
(\ref{18}) given in the text in order to 
solve the internal bubble, we find
\begin{eqnarray}\label{A5}
& I_{3}^{(\tau)} (k,i=0;\sigma) = \sigma  \frac{1}{\epsilon}
\Bigl((1-\frac{\epsilon}{2})\int_{0}^{1} dx \sum_{j=-\infty}^{\infty} 
\int \frac{d^{d-1}q}{[(q)^{2} + \sigma^{2}(j+\tau)^{2}+1][x(1-x)[(q+k)^{2} 
+ \sigma^{2}j^{2}] + 1]^{\frac{\epsilon}{2}}} \nonumber\\
& + \frac{\epsilon}{2} \Gamma(2-\frac{\epsilon}{2}) 
\int \frac{d^{d-1}q}{[(q)^{2} + \sigma^{2}(j+\tau)^{2}+1]} F^{(\tau)}_{\frac{\epsilon}{2}}(q+k,i=0;\sigma)\Bigr).
\end{eqnarray} 
\par Before going ahead define the objects
\begin{equation}\label{A6}
F_{\alpha,\beta}^{(\tau)}(k,i;\sigma) \equiv \frac{1}{S_{d}}\sigma \Sigma_{j=-\infty}^{\infty} 
\int d^{d-1}q \frac{F_{\alpha}^{(\tau)}(q+k,j+i;\sigma)}{[(q)^{2} + \sigma^{2}(j+\tau)^{2}+ 1]^{\beta}}, 
\end{equation}
\begin{equation}\label{A7}
F_{\alpha}^{\prime (\tau)}(\sigma) \equiv \frac{\partial F_{\alpha,1}^{(\tau)}(k,i;r)}
{\partial k^{2}}\big|_{(k,i)=0},
\end{equation}\label{A8}
\begin{equation}
i_{3}(k,\sigma,x)= \sum_{j=-\infty}^{\infty} 
\int \frac{d^{d-1}q}{[q^{2} + \sigma^{2}(j+\tau)^{2}+1][(q+k)^{2} 
+ \sigma^{2} j^{2} + \frac{1}{x(1-x)}]^{\frac{\epsilon}{2}}},
\end{equation}
which shall be important in what follows. Inserting last equations 
into the expression for $I_{3}^{(\tau)}$, the latter can be rewritten as 
\begin{eqnarray}\label{A9}
I_{3}^{(\tau)} (k,i=0;\sigma) &=& \frac{1}{\epsilon}
\Bigl(\sigma(1-\frac{\epsilon}{2})\int_{0}^{1} dx [x(1-x)]^{-\frac{\epsilon}{2}} 
i_{3}(k,\sigma,x) \nonumber\\
&& + S_{d} \frac{\epsilon}{2} \Gamma(2-\frac{\epsilon}{2})
F^{(\tau)}_{\frac{\epsilon}{2},1}(k,i=0;\sigma)\Bigr).
\end{eqnarray} 
Next, let us compute $i_{3}(k,\sigma,x)$. First, perform the change of 
variables $q'=q+k$. We utilize another Feynman parameter 
which leads to
\begin{eqnarray}\label{A10}
&&i_{3}(k,\sigma,x)= \frac{\Gamma(1+\frac{\epsilon}{2})}
{\Gamma(\frac{\epsilon}{2})}\int_{0}^{1} dy (1-y)^{\frac{\epsilon}{2}-1} 
\sum_{j=-\infty}^{\infty} 
\int d^{d-1}q' \nonumber\\
&& \times
\frac{1}{\Bigl(q^{' 2} - 2ykq' + yk^{2}+ y + \sigma^{2}(j+y \tau)^{2} 
+ y(1-y) \sigma^{2}\tau^{2} + \frac{1-y}{x(1-x)}\Bigr)^{1+\frac{\epsilon}{2}}} \;.
\end{eqnarray}
Resolving the momentum integral using Eq.(\ref{10}), we obtain after expanding in $\epsilon=4-d$ the result
\begin{eqnarray}\label{A11}
&&i_{3}(k,\sigma,x)= \frac{S_{d-1} \Gamma(\frac{d-1}{2}) 
\Gamma(-\frac{1}{2} + \epsilon)}
{2 \Gamma(\frac{\epsilon}{2})}\int_{0}^{1} dy (1-y)^{\frac{\epsilon}{2}-1}
\nonumber\\ 
&&\sum_{j=-\infty}^{\infty} \Bigr[y(1-y)[k^{2}+ \sigma^{2} \tau^{2}] + 
\sigma^{2} (j+ y \tau)^{2}+ y + \frac{1-y}{x(1-x)}\Bigr]^{\frac{1}{2}-\epsilon} \;.
\end{eqnarray}
Utilize the identity $\sqrt{\pi}\Gamma(\frac{d-1}{2})S_{d-1}= \Gamma(\frac{d}{2}) S_{d}$, absorb $S_{d}$ in the redefinition of the coupling constant just 
as explained for the one-loop four-point graph in the main text and expand in 
$\epsilon$. Pluging these steps in $I_{3}^{(\tau)} (k,i=0;\sigma)$, it is 
easy to show that 
\begin{eqnarray}\label{A12}
&&I_{3}^{(\tau)}(k,i=0,\sigma)= \frac{1}{\epsilon}\Bigl[\sigma (1-\frac{\epsilon}{2})\frac{\Gamma(2-\frac{\epsilon}{2})\Gamma(-\frac{1}{2} + \epsilon)}{2\sqrt{\pi} \Gamma(\frac{\epsilon}{2})} 
\int_{0}^{1} dx [x(1-x)]^{-\frac{\epsilon}{2}} \nonumber\\ 
&& \times \; \int_{0}^{1} dy (1-y)^{\frac{\epsilon}{2}-1} \sum_{j=-\infty}^{\infty} \Bigl[y(1-y)[k^{2}+ \sigma^{2} \tau^{2}] + 
\sigma^{2} (j+ y \tau)^{2}+ y + \frac{1-y}{x(1-x)}\Bigr]^{\frac{1}{2}-\epsilon}
\nonumber\\
&& \;+ \;\; \frac{\epsilon}{2} \Gamma(2-\frac{\epsilon}{2})
F^{(\tau)}_{\frac{\epsilon}{2},1}(k,i=0;\sigma)\Bigr].
\end{eqnarray}
Now, perform the derivative with respect to $k^{2}$ at $k^{2}=0$. Define 
$I_{3SP}^{\prime (\tau)}= \frac{\partial I_{3}^{(\tau)}(k,i=0,\sigma)}
{\partial k^{2}}|_{k^{2}=0}$, expanding the argument of the $\Gamma$-functions in 
$\epsilon(=4-d)$, employing the representations Eqs.(\ref{12}) and 
(\ref{13}) to resolving the remaining summation and neglecting higher order 
terms in $\epsilon$, we get to
\begin{eqnarray}\label{A13}
&&I_{3SP}^{\prime (\tau)}(\sigma)= - \frac{1}{4\epsilon} 
\Bigl[(1- \epsilon) 
\int_{0}^{1} dx [x(1-x)]^{-\frac{\epsilon}{2}} 
\int_{0}^{1} dy y (1-y)^{\frac{\epsilon}{2}}\Bigl[y(1-y)\tau^{2} + y\sigma^{-2} \nonumber\\
&& + \;\frac{1-y}{x(1-x)\sigma^{2}}\Bigr]^{-\epsilon} + \;\epsilon \int_{0}^{1} dx [x(1-x)]^{-\frac{\epsilon}{2}} 
\int_{0}^{1} dy y (1-y)^{\frac{\epsilon}{2}}\nonumber\\
&& \times\; 
f_{\frac{1}{2} + \epsilon} \Bigl(y \tau,\sqrt{y(1-y) \tau^{2} + \sigma^{-2} y + 
\frac{1-y}{x(1-x)} \sigma^{-2}}\Bigr) 
- 2 \epsilon F^{\prime}_{\frac{\epsilon}{2}}(\sigma) \Bigr].
\end{eqnarray}
This is the explicit form that can be reduced further by noticing that 
$O(\epsilon)$ terms in this expression can be computed at $\epsilon=0$. We can 
simplify the final steps when we write last equation in terms of 
the following parametric integrals
\begin{subequations}\label{A14}
\begin{eqnarray}
&G^{(\tau)}(r) = -2 \int dx dy y \;\;\; \times\nonumber\\ 
& ln\Bigl[y(1-y) \tau^{2} r^{2} + y + \frac{1-y}{x(1-x)}\Bigr] -\frac{1}{2},\label{A14a}\\
&H^{(\tau)}(r) = 2 \int dx dy  y \;\;\; \times \nonumber\\ 
&f_{\frac{1}{2}}\Bigl(y\tau,\sqrt{y(1-y)\tau^{2} + r^{-2} y + \frac{r^{-2}(1-y)}{x(1-x)}}\Bigr),\label{A14b}
\end{eqnarray}
\end{subequations}
with $r \equiv \sigma$. Inserting these definitions 
in the remainder of last integral we finally obtain  
\begin{equation}\label{A15}
I_{3SP}^{\prime (\tau)} (\sigma)= -\frac{1}{8\epsilon}
\bigl(1-\frac{\epsilon}{4} + \epsilon W^{(\tau)}(\sigma)\bigr), 
\end{equation}
where $W^{(\tau)}(r)= G^{(\tau)}(\sigma) + H^{(\tau)}(\sigma) -
4 F_{0}^{\prime (\tau)}(\sigma)$. Let us compute the three-loop 
contribution for the two-point function following the same reasoning. We can 
write that integral in the form
\begin{eqnarray}\label{A16}
&& I^{(\tau)}_{5}(k,i=0;\sigma) = \sigma \sum_{j=-\infty}^{\infty} 
\int d^{d-1}q \frac{[I_{2}^{(\tau)}(q,j;\sigma)]^{2}}{[(q-k)^{2} + \sigma^{2}(j+\tau)^{2}
+1]}. 
\end{eqnarray} 
Making use of the result for the one-loop bubble and neglecting higher order 
corrections in $\epsilon$ we obtain the following intermediary result 
\begin{eqnarray}\label{A17}
&& I^{(\tau)}_{5}(k,i=0;\sigma) = \sigma \sum_{j=-\infty}^{\infty} \int 
\frac{d^{d-1}q}{[(q-k)^{2} + \sigma^{2}(j+\tau)^{2}+1]} \frac{1}{\epsilon^{2}}
\Bigl[(1-\epsilon) \times \nonumber\\
&& \Bigl[\int_{0}^{1} \frac{dx}{[x(1-x)q^{2}+ j^{2} \tau^{2} + 1]^{\frac{\epsilon}{2}}}\Bigr]^{2} 
+ \epsilon F_{\frac{\epsilon}{2}}^{(\tau)}(q,\sigma) \int_{0}^{1} dx 
[x(1-x)q^{2}+ j^{2} \tau^{2} + 1]^{\frac{-\epsilon}{2}}\Bigr].
\end{eqnarray}
It is not difficult to prove that we can rewrite last equation as
\begin{eqnarray}\label{A18}
&& I^{(\tau)}_{5}(k,i=0;\sigma) = \frac{1}{\epsilon^{2}}
\Bigl[\sigma (1+\epsilon)\sum_{j=-\infty}^{\infty} \int 
\frac{d^{d-1}q}{[(q-k)^{2} + \sigma^{2}(j+\tau)^{2}+1]}\nonumber\\
&& \times \int_{0}^{1}\frac{dx}{[q^{2} + \sigma^{2} j^{2} 
+ \frac{1}{x(1-x)}]^{\epsilon}} + S_{d} 
\epsilon F_{\frac{\epsilon}{2},1}^{(\tau)}(k,\sigma)+ O(\epsilon^{2})\Bigr].
\end{eqnarray}
Define the subintegral
\begin{equation}\label{A19}
i_{5}(k,\sigma,x)= \sum_{j=-\infty}^{\infty} 
\int \frac{d^{d-1}q}{[(q-k)^{2} + \sigma^{2}(j+\tau)^{2}+1][q^{2} 
+ \sigma^{2} j^{2} + \frac{1}{x(1-x)}]^{\epsilon}}.
\end{equation}
We utilize another Feynman parameter in 
last integral in order to compute the momentum integral 
employing Eq.(\ref{10}). The result in terms of parametric integrals reads
\begin{eqnarray}\label{A20}
&& i_{5}(k,\sigma,x) = \frac{S_{d-1} \Gamma(\frac{d-1}{2}) 
\Gamma(-\frac{1}{2} + \frac{3\epsilon}{2})}
{2 \Gamma(\epsilon)}\int_{0}^{1} dy (1-y)^{\epsilon - 1}
\nonumber\\ 
&&\sum_{j=-\infty}^{\infty} \Bigr[y(1-y)[k^{2}+ \sigma^{2} \tau^{2}] + 
\sigma^{2} (j+ y \tau)^{2}+ y + \frac{1-y}{x(1-x)}\Bigr]^{\frac{1}{2}-\frac{3\epsilon}{2}} \;.
\end{eqnarray} 
Once again, use the identity $\sqrt{\pi}\Gamma(\frac{d-1}{2})S_{d-1}= \Gamma(\frac{d}{2}) S_{d}$ and expand in $\epsilon$. Replacing it back into 
$I_{5}^{(\tau)}(k,\sigma)$ and absorbing the factor $S_{d}$, we find
\begin{eqnarray}\label{A21}
&&I_{5}^{(\tau)}(k,i=0,\sigma)=  \frac{1}{\epsilon^{2}}\Bigl[\sigma \frac{\Gamma(2-\frac{\epsilon}{2}) 
\Gamma(-\frac{1}{2} + \frac{3\epsilon}{2})}
{2\sqrt{\pi} \Gamma(\epsilon)}(1+\epsilon) 
\int_{0}^{1} dx  \nonumber\\ 
&& \times \; \int_{0}^{1} dy (1-y)^{\epsilon-1} \sum_{j=-\infty}^{\infty} \Bigl[y(1-y)[k^{2}+ \sigma^{2} \tau^{2}] + 
\sigma^{2} (j+ y \tau)^{2}+ y + \frac{1-y}{x(1-x)}\Bigr]^{\frac{1}{2}-\frac{3\epsilon}{2}}
\nonumber\\
&& \;+ \;\; \epsilon
F^{(\tau)}_{\frac{\epsilon}{2},1}(k,i=0;\sigma)\Bigr].
\end{eqnarray}
Now performing the derivative and defining $I_{5SP}^{\prime (\tau)}= \frac{\partial I_{5}^{(\tau)}(k,i=0,\sigma)}
{\partial k^{2}}|_{k^{2}=0}$ results in the expression
\begin{eqnarray}\label{A22}
&& I_{5SP}^{\prime (\tau)}(\sigma)= \frac{1}{\epsilon^{2}}\Bigl[\sigma \frac{\Gamma(2-\frac{\epsilon}{2}) 
\Gamma(-\frac{1}{2} + \frac{3\epsilon}{2})}
{2\sqrt{\pi} \Gamma(\epsilon)}(1+\epsilon)(\frac{1}{2}-\frac{3\epsilon}{2}) 
\int_{0}^{1} dx  \nonumber\\ 
&& \times \int_{0}^{1} dy y (1-y)^{\epsilon} 
\sum_{j=-\infty}^{\infty} \Bigl[y(1-y) \sigma^{2} \tau^{2} 
+ \sigma^{2} (j+ y \tau)^{2}+ y + \frac{1-y}{x(1-x)}\Bigr]^{-\frac{1}{2}-\frac{3\epsilon}{2}}
\nonumber\\
&& \;+ \;\; \epsilon
F^{\prime (\tau)}_{\frac{\epsilon}{2}}(\sigma)\Bigr].
\end{eqnarray}
Performing the summation we can rewrite this integral as 
\begin{eqnarray}\label{A23}
&& I_{5SP}^{\prime (\tau)}(\sigma)= \frac{1}{\epsilon^{2}}\Bigl[\sigma \frac{\Gamma(2-\frac{\epsilon}{2}) 
\Gamma(-\frac{1}{2} + \frac{3\epsilon}{2})}
{2 \Gamma(\frac{1}{2}+\frac{3\epsilon}{2}) \Gamma(\epsilon)}(1+\epsilon)(\frac{1}{2}-\frac{3\epsilon}{2}) 
\int_{0}^{1} dx \int_{0}^{1} dy y (1-y)^{\epsilon}\nonumber\\
&& \times \Bigr[\Gamma(\frac{3\epsilon}{2}) 
\Bigl[y(1-y)\tau^{2} + y \sigma^{-2} 
+ \frac{1-y}{x(1-x)\sigma^{2}}\Bigr]^{-\frac{3 \epsilon}{2}}\nonumber\\
&& + f_{\frac{1}{2}+\frac{3 \epsilon}{2}}\Bigl(y\tau,\sqrt{y(1-y)\tau^{2} 
+ y \sigma^{-2} + \frac{1-y}{x(1-x)\sigma^{2}}}\Bigr)\Bigr]  
\;+ \epsilon F^{\prime (\tau)}_{\frac{\epsilon}{2}}(\sigma)\Bigr].
\end{eqnarray}
Expanding the $\Gamma$-functions and setting $\epsilon=0$ in the subscripts 
of the functions which remain just as we proceeded in the calculation of the 
two-loop contribution we obtain the result
\begin{equation}\label{A24}
I_{5}^{\prime (\tau)} (\sigma)= -\frac{1}{6\epsilon^{2}}
\bigl(1-\frac{\epsilon}{4} + \frac{3 \epsilon}{2} W^{(\tau)}(\sigma)\bigr),
\end{equation} 
\par The integral $I_{4SP}(\sigma)$ can be solved along similar steps. 
Realizing that, it can be written at the symmetry point in terms of the 
four-point one-loop subdiagram as 
\begin{equation}\label{A25}
I_{4SP}^{(\tau)}(\sigma) = \sigma \sum_{j=-\infty}^{\infty} 
\int d^{d-1}q 
\frac{I_{2}^{(\tau)}(q,j;\sigma)}{[(q)^{2} + \sigma^{2}(j+\tau)^{2}+ 1]^{2}}, 
\end{equation}  
we can solve the internal bubble, and obtain
\begin{eqnarray}\label{A26}
&& I_{4SP}^{(\tau)}(\sigma) = [\sigma\frac{1}{\epsilon}(1-\frac{\epsilon}{2})
\int_{0}^{1} dx \sum_{j=-\infty}^{\infty} \int d^{d-1}q 
\frac{1}{[(q)^{2} + \sigma^{2}(j+\tau)^{2}+ 1]^{2}}\nonumber\\
&&\;\times \frac{1}{[(q^{2} + \sigma^{2}\tau^{2})x(1-x) + 1]^{\frac{\epsilon}{2}}} 
+ \frac{1}{2}F_{\frac{\epsilon}{2}, 2}^{(\tau)}(\sigma) S_{d}].
\end{eqnarray}
Before proceeding, let us prove that last term is convergent, does not 
contribute to the ultraviolet divergences of $I_{4SP}^{(\tau)}(\sigma)$ and 
therefore can be neglected in the consideration of its singularities 
(dimensional poles in $\epsilon$). Explicitly, it is given by Eq.(\ref{A6})
\begin{equation}\label{A27}
F_{\frac{\epsilon}{2}, 2}^{(\tau)}(\sigma) \equiv \frac{1}{S_{d}}\sigma \Sigma_{j=-\infty}^{\infty} 
\int d^{d-1}q \frac{F_{\frac{\epsilon}{2}}^{(\tau)}(q,j;\sigma)}{[q^{2} + \sigma^{2}(j+\tau)^{2}+ 1]^{2}}. 
\end{equation}
Since this is a massive integral, the potential singularities come from the 
region of high momentum. From simple power counting together with our previous 
dicussions, this integral will be 
divergent if, in the limit $q \rightarrow \infty$, the  
$F_{\frac{\epsilon}{2}}^{(\tau)}(q,j;\sigma)$ behavior is $O(q^{0})$ or 
proportional to a positive power of $q$. Thus, it suffices to prove that this 
object is proportional to a negative power of $q$, which we shall show next. 
From the definitions given in the main text Eq.(\ref{17}) and the 
representation in terms of sum involving the product of cosine and Bessel 
function, we can write it as
\begin{eqnarray}\label{A28}
&& F^{(\tau)}_{\frac{\epsilon}{2}}(q,i;\sigma)= 4 \sigma^{-\epsilon} 
\int_{0}^{1} dx \sum_{m=1}^{\infty} cos[2\pi m(\tau+ix)]\Bigl(\frac{\pi m}
{\sigma^{-1} \sqrt{x(1-x)(q^{2} + \sigma^{2} i^{2}) +1}}\Bigr)^{\frac{\epsilon}{2}} \nonumber\\
&& \;\;\times\;K_{\frac{\epsilon}{2}}\Bigl(2\pi m \sigma^{-1} \sqrt{x(1-x)(q^{2} + \sigma^{2} i^{2}) +1}\Bigr).
\end{eqnarray}
In order to attain maximal simplicity, take $i=0$ in the above expression, 
since it is obvious that $ F^{(\tau)}_{\frac{\epsilon}{2}}(q,i;\sigma)< 
F^{(\tau)}_{\frac{\epsilon}{2}}(q,i=0;\sigma)$. Note that the integrand is 
symmetric around $x=\frac{1}{2}$ and we can write
\begin{eqnarray}\label{A29}
&& F^{(\tau)}_{\frac{\epsilon}{2}}(q,i=0;\sigma)= 8 \sigma^{-\epsilon} 
\int_{0}^{\frac{1}{2}} dx \sum_{m=1}^{\infty} cos(2\pi m\tau)\Bigl(\frac{\pi m}
{\sigma^{-1} \sqrt{x(1-x)q^{2}+1}}\Bigr)^{\frac{\epsilon}{2}} \nonumber\\
&& \;\;\times\;K_{\frac{\epsilon}{2}}\Bigl(2\pi m \sigma^{-1} 
\sqrt{x(1-x)q^{2}+1}\Bigr).
\end{eqnarray}
In the limit $q \rightarrow \infty$, choose a small real parameter 
$\lambda<<1$ with the property  $\lambda q^{2} \rightarrow \infty$. The idea 
is to split the integration limits into two pieces: in the first one we use 
the Bessel function and in the second piece we replace its asymptotic form 
for large values of the argument, namely
\begin{eqnarray}\label{A30}
&& \underset{q \rightarrow \infty}{lim}F^{(\tau)}_{\frac{\epsilon}{2}}(q,i=0;\sigma)= 8 \sigma^{-\epsilon} 
\sum_{m=1}^{\infty}\Bigl[\int_{0}^{\lambda} dx  cos(2\pi m\tau)
\Bigl(\frac{\pi m}
{\sigma^{-1} \sqrt{xq^{2}+1}}\Bigr)^{\frac{\epsilon}{2}} \nonumber\\
&& \;\;\times\;K_{\frac{\epsilon}{2}}\Bigl(2\pi m \sigma^{-1} 
\sqrt{xq^{2}+1}\Bigr) + \int_{\lambda}^{\frac{1}{2}} dx  cos(2\pi m\tau)
\Bigl(\frac{\pi m}
{\sigma^{-1} \sqrt{x(1-x)q^{2}+1}}\Bigr)^{\frac{\epsilon}{2}}\times \nonumber\\
&& \sqrt{\frac{\pi}{4 \pi m \sigma^{-1}\sqrt{x(1-x)q^{2}+1}}}exp\Bigl(-2\pi m 
\sigma^{-1}\sqrt{x(1-x)q^{2}+1}\Bigr)\Bigr].
\end{eqnarray}
The second term can be neglected in this limit. Performing the change of 
variables $y=1+xq^{2}$, we can rewrite last expression taking into account 
these observations in the form
\begin{eqnarray}\label{A31}
&& \underset{q \rightarrow \infty}{lim}F^{(\tau)}_{\frac{\epsilon}{2}}(q,i=0;\sigma)= \frac{8\sigma^{-\frac{\epsilon}{2}}}{q^{2}} 
\sum_{m=1}^{\infty} cos(2\pi m\tau)(\pi m)^{\frac{\epsilon}{2}} 
\int_{1}^{\infty} dy  y^{-\frac{\epsilon}{4}}\nonumber\\
&& \;\times\; K_{\frac{\epsilon}{2}}(2\pi m \sigma^{-1}y).
\end{eqnarray}
Using the identity \cite{GR}
\begin{equation}\label{A32}
\int_{1}^{\infty} dx x^{-\frac{\nu}{2}}(x-1)^{\mu-1}K_{\nu}(a\sqrt{x}) = 
\Gamma(\mu)2^{\mu}a^{-\mu}K_{\nu-\mu}(a),
\end{equation}
we are able to prove that 
\begin{equation}\label{A33}
\underset{q \rightarrow \infty}{lim}F^{(\tau)}_{\frac{\epsilon}{2}}(q,i=0;\sigma)= \frac{8\sigma^{1-\frac{\epsilon}{2}}}{q^{2}} 
\sum_{m=1}^{\infty} cos(2\pi m\tau)(\pi m)^{\frac{\epsilon}{2}-1} 
 K_{\frac{\epsilon}{2}-1}(2\pi m \sigma^{-1}),
\end{equation}
which is clearly regular in $\epsilon$ completing our task in 
proving that in the ultraviolet region this object and therefore the 
desired integral involving it are both finite. Then it is safe to neglect that 
term in the computation of $I_{4SP}^{(\tau)}(\sigma)$.
\par When we neglect the correction from Eq.(\ref{A26}) as 
just discussed, we can proceed from the following expression
\begin{eqnarray}\label{A34}
&& I_{4SP}^{(\tau)}(\sigma) = \sigma\frac{1}{\epsilon}(1-\frac{\epsilon}{2})
\int_{0}^{1} dx[x(1-x)]^{\frac{-\epsilon}{2}} \sum_{j=-\infty}^{\infty} 
\int d^{d-1}q \frac{1}{[q^{2} + \sigma^{2}(j+\tau)^{2}+ 1]^{2}}\nonumber\\
&&\;\times \frac{1}{[(q^{2} + \sigma^{2}\tau^{2}) + \frac{1}{x(1-x)}]^{\frac{\epsilon}{2}}}.
\end{eqnarray}
\par We introduce an additional Feynman parameter $y$ to rewrite last equation 
as
\begin{eqnarray}\label{A35}
&& I_{4SP}^{(\tau)}(\sigma) = \sigma 
\frac{\Gamma(2+\frac{\epsilon}{2})}{\Gamma(\frac{\epsilon}{2})\epsilon}(1-\frac{\epsilon}{2})
\int_{0}^{1} dx[x(1-x)]^{\frac{-\epsilon}{2}}\sum_{j=-\infty}^{\infty} 
\int_{0}^{1} dy y(1-y)^{\frac{\epsilon}{2}-1}\nonumber\\
&& \times \int \frac{d^{d-1}q}{\Bigl[q^{2}+\sigma^{2}(j+y\tau)^{2} 
+ y(1-y)\sigma^{2}\tau^{2} + y 
+ \frac{1-y}{x(1-x)}\Bigr]^{2+\frac{\epsilon}{2}}}.
\end{eqnarray}
Utilizing Eq.(\ref{10}), we resolve the momentum integral and express the 
result in the form
\begin{eqnarray}\label{A36}
&& I_{4SP}^{(\tau)}(\sigma) = \sigma 
\frac{\Gamma(1+\frac{\epsilon}{2})S_{d-1}\Gamma(\frac{d-1}{2})}{2\Gamma(\frac{\epsilon}{2})\epsilon}(1-\frac{\epsilon}{2})
\int_{0}^{1} dx[x(1-x)]^{\frac{-\epsilon}{2}} 
\int_{0}^{1} dy y(1-y)^{\frac{\epsilon}{2}-1}\nonumber\\ 
&& \sum_{j=-\infty}^{\infty}\Bigl[\sigma^{2}(j+y\tau)^{2} 
+ y(1-y)\sigma^{2}\tau^{2} + y 
+ \frac{1-y}{x(1-x)}\Bigr]^{-\frac{1}{2}-\epsilon}.
\end{eqnarray}
Performing the summation using Eq.(\ref{12}), employing the identity 
$\sqrt{\pi}\Gamma(\frac{d-1}{2})S_{d-1}= \Gamma(\frac{d}{2}) S_{d}$, expanding
the argument of the $\Gamma$-function in $\epsilon$ and absorbing the factor 
$S_{d}$, we obtain the following result
\begin{eqnarray}\label{A37}
&& I_{4SP}^{(\tau)}(\sigma) = \frac{1}{\epsilon} \Bigl[(1-\frac{\epsilon}{2})
\sigma^{-2\epsilon} \frac{\Gamma(2-\frac{\epsilon}{2})}
{2\Gamma(\frac{\epsilon}{2})}\int_{0}^{1} dx[x(1-x)]^{\frac{-\epsilon}{2}}
\int_{0}^{1} dy y(1-y)^{\frac{\epsilon}{2}-1}\nonumber\\
&&\;\;\times \Bigl[\Gamma(\epsilon)\Bigl(y(1-y)\tau^{2} + y\sigma^{-2}
+ \frac{(1-y)\sigma^{-2}}{x(1-x)}\Bigr)^{-\epsilon} \nonumber\\
&& \;\; + f_{\frac{1}{2}+ \epsilon}\Bigr(y\tau,\sqrt{y(1-y)\tau^{2} + y\sigma^{-2} 
+ \frac{(1-y)\sigma^{-2}}{x(1-x)}} \Bigr) \Bigr] 
+ \frac{\epsilon}{2}F_{\frac{\epsilon}{2},2}^{(\tau)}(\sigma)\Bigr].
\end{eqnarray}
Note that the integral over $y$ possesses a pole in $y=1$. A standard 
procedure is to compute the bracket which multiplies the integral at 
$y=1$ \cite{amit}, which facilitates the computation and retains the pole 
contribution which we are interested. Expanding in 
$\epsilon$ and neglecting $O(\epsilon^{0})$ terms, we finally obtain
\begin{equation}\label{A38}
I_{4SP}^{(\tau)} (\sigma)= \frac{1}{2 \epsilon^{2}}
\bigl((1-\frac{\epsilon}{2}) + \epsilon f_{\frac{1}{2}}(\tau,\sigma^{-1})\bigr).
\end{equation}
\par These integrals and all the nomenclature defined here are utilized in 
Secs. II and III.   

\section{Massless integrals in dimensional regularization}
\par The higher-loop massless integrals are even simpler to evaluate than 
those occurring in the massive setting discussed above, with the difference 
that they need to be calculated at nonvanishing external momenta. The massless 
counterparts of the integrals discussed in the previous Appendix are given by 
\begin{eqnarray}\label{B1}
&& I_{3}(k,i,\sigma) = \sigma^{2}\sum_{j_{1},j_{2}=-\infty}^{\infty} 
\int \frac{d^{d-1}{q_{1}}d^{d-1}q_{2}}
{\left( q_{1}^{2} + \sigma^{2}(j_{1}+\tau)^2 \right)
\left( q_{2}^{2} + \sigma^{2}(j_{2}+\tau)^2 \right)} \nonumber\\
&& \times \;\;\; \frac{1}{[(q_{1} + q_{2} + k)^{2} 
+ \sigma^{2}(j_{1} + j_{2}+ i + \tau)^{2} ]}\;\;,
\end{eqnarray}
\begin{eqnarray}\label{B2}
&&I_{4}(k_{1},k_{2},k_{3}, k_{4},i_{1},i_{2},i_{3}, i_{4},\sigma) = \sigma^{2}\sum_{j_{1},j_{2}=-\infty}^{\infty}
\int \frac{d^{d-1}{q_{1}}d^{d-1}q_{2}}
{\left(q_{1}^{2} + \sigma^{2}(j_{1} + \tau)^2 \right)} \nonumber\\
&& \;\times \frac{1}{\left(q_{2}^{2} + \sigma^{2}(j_{2} + \tau)^2 \right)[(q_{1} - q_{2} + k_{3})^{2} + \sigma^{2}(j_{1} - j_{2} + i_{3} + \tau)^2]} \nonumber\\
&&\;\;\times \frac{1}{[(P - q_{1})^{2} + \sigma^{2}(p - j_{1} 
+ \tau)^2]}\;\;,
\end{eqnarray}
\begin{eqnarray}\label{B3}
&& I_{5}(k,i,\sigma) = \sigma^{2}\sum_{j_{1},j_{2}=-\infty}^{\infty}
\int \frac{d^{d-1}{q_{1}}d^{d-1}q_{2}d^{d-1}q_{3}}
{\left(q_{1}^{2} + \sigma^{2}(j_{1}+\tau)^{2} \right) 
\left( q_{2}^{2} +  \sigma^{2}(j_{2}+\tau)^{2} \right)} \nonumber\\
&& \times \frac{1}{\left( q_{3}^{2} +  \sigma^{2}(j_{3}+\tau)^{2} 
+ 1 \right)[ (q_{1} + q_{2} + k)^{2} 
+ \sigma^{2}(j_{1}+j_{2}+i+\tau)^{2}]} \nonumber\\
&& \times \frac{1}{[(q_{1} + q_{3} + k)^{2} 
+ \sigma^{2}(j_{1}+j_{3}+i+\tau)^{2}]}.
\end{eqnarray}
\par Firstly, let us discuss $I_{3}$ at arbitrary external momentum, which 
shall be important in the evaluation of critical exponents using minimal 
subtraction. As we are going to compute the exponents in normalization 
conditions as well, we shall discuss its derivative with respect to squared 
nonvanishing external momenta for that sake. All integrals will be computed at 
zero external quasimomenta from now on. Since many developments discussed 
in the previous Appendix will have their exact analogy here, we will be more 
economical in the steps to compute the integrals. First, write $I_{3}$ in 
terms of the one-loop subdiagram as
\begin{eqnarray}\label{B4}
&& I^{(\tau)}_{3}(k,i=0;\sigma) = \sigma \sum_{j=-\infty}^{\infty} 
\int d^{d-1}q \frac{I_{2}^{(\tau)}(q,j;\sigma)}{[(q-k)^{2} + \sigma^{2}(j+\tau)^{2}]}. 
\end{eqnarray} 
Replacing the value of the subdiagram computed in Sec. IV 
through the use of Eqs.(\ref{48})-(\ref{50}) defined for the {\it massless} 
case, we get 
\begin{eqnarray}\label{B5}
& I_{3}^{(\tau)} (k,i=0;\sigma) =  \frac{1}{\epsilon}
\Bigl(\sigma (1-\frac{\epsilon}{2}) 
\int_{0}^{1} dx [x(1-x)]^{-\frac{\epsilon}{2}}
\sum_{j=-\infty}^{\infty} 
\int \frac{d^{d-1}q}{[(q-k)^{2} + \sigma^{2}(j+\tau)^{2}]
[q^{2} + \sigma^{2}j^{2}]^{\frac{\epsilon}{2}}} \nonumber\\
& + \frac{\epsilon}{2} \Gamma(2-\frac{\epsilon}{2}) \sigma
\int \frac{d^{d-1}q}{[(q-k)^{2} + \sigma^{2}(j+\tau)^{2}+1]} F^{(\tau)}_{\frac{\epsilon}{2}}(q,i=0;\sigma)\Bigr).
\end{eqnarray} 
\par For the sake of minimal subtraction, define the parametric integral 
\begin{equation}\label{B6}
L_{3}(k^{2}+\sigma^{2}\tau^{2})=\int_{0}^{1}dyyln[y(1-y)(k^{2}+\sigma^{2}\tau^{2})].
\end{equation} 
Other useful definitions are the massless functions
\begin{subequations}\label{B7}
\begin{eqnarray}
&& F_{\alpha,\beta}^{(\tau)}(k,i;\sigma) \equiv \frac{1}{S_{d}}\sigma \sum_{j=-\infty}^{\infty} 
\int d^{d-1}q \frac{F_{\alpha}^{(\tau)}(q+k,j+i;\sigma)}{[q^{2} 
+ r^{2}(j+\tau)^{2}]^{\beta}}, \label{B7a}\\
&& F_{\alpha}^{\prime (\tau)}(k,i;\sigma) \equiv 
\frac{\partial F_{\alpha,1}^{(\tau)}(k,i;\sigma)}
{\partial k^{2}}\big|_{(k^{2}=1,i=0)},\label{B7b}
\end{eqnarray}
\end{subequations}
\begin{equation}\label{B8}
i_{3}(k,\sigma)= \sum_{j=-\infty}^{\infty} 
\int \frac{d^{d-1}q}{[q^{2} + \sigma^{2}(j+\tau)^{2}][(q+k)^{2} 
+ \sigma^{2} j^{2}]^{\frac{\epsilon}{2}}}.
\end{equation}
\par Let us compute $i_{3}(k,\sigma)$, which is analogous to the expression 
for the massive case and much simpler. We introduce a Feynman parameter $y$ 
in order to express the two denominators as a single one, solve the 
momentum integral using Eq.(\ref{10}) and expand in $\epsilon$ afterward. 
This set of steps are identical to those which led to Eq.(\ref{A11}) 
and the reader can check that it results in the following expression
\begin{eqnarray}\label{B9}
&&i_{3}(k,\sigma)= \frac{S_{d-1} \Gamma(\frac{d-1}{2}) 
\Gamma(-\frac{1}{2} + \epsilon)}
{2 \Gamma(\frac{\epsilon}{2})}\int_{0}^{1} dy (1-y)^{\frac{\epsilon}{2}-1}
\nonumber\\ 
&&\sum_{j=-\infty}^{\infty} \Bigr[y(1-y)(k^{2} + \sigma^{2}\tau^{2}) 
+ \sigma^{2} (j+ y \tau)^{2}\Bigr]^{\frac{1}{2}-\epsilon} \;.
\end{eqnarray}
The summation can be performed as before and we find
\begin{eqnarray}\label{B10}
&&i_{3}(k,\sigma)= \frac{S_{d-1} \Gamma(\frac{d-1}{2})}
{2 \Gamma(\frac{\epsilon}{2})}\int_{0}^{1} dy (1-y)^{\frac{\epsilon}{2}-1}
\sqrt{\pi} \Bigr[\sigma^{-1}\Gamma(\epsilon-1)\Bigr(y(1-y)[k^{2}\nonumber\\ 
&& + \sigma^{2}\tau^{2}]\Bigl)^{1-\epsilon} + \sigma^{1-2\epsilon}f_{-\frac{1}{2} + \epsilon}(y \tau,\sigma^{-1}\sqrt{y(1-y)[k^{2} + \sigma^{2}\tau^{2}]})\Bigl].
\end{eqnarray}
\par Utilize the identity 
$\sqrt{\pi}\Gamma(\frac{d-1}{2})S_{d-1}= \Gamma(\frac{d}{2}) S_{d}$, 
absorb $S_{d}$ in the redefinition of the coupling constant just 
as explained for the one-loop four-point graph in the main text and 
expand in $\epsilon$. Integrating over the Feynman parameter $x$ and 
substituting into the expression of $I_{3}^{(\tau)}(k,i=0,\sigma)$, we 
obtain 
\begin{eqnarray}\label{B11}
&& I_{3}^{(\tau)} (k,i=0;\sigma) =  \frac{1}{\epsilon}
\Bigl((1-\frac{\epsilon}{2}) 
\frac{\Gamma^{2}(1-\frac{\epsilon}{2}) \Gamma(2-\frac{\epsilon}{2})}
{2\Gamma(\frac{\epsilon}{2})\Gamma(2-\epsilon)}\int_{0}^{1} dy (1-y)^{\frac{\epsilon}{2}-1}
\Bigr[\Gamma(\epsilon-1)\nonumber\\ 
&& \times \Bigr(y(1-y)[k^{2} 
+ \sigma^{2}\tau^{2}]\Bigl)^{1-\epsilon} +  \sigma^{2-2\epsilon}f_{-\frac{1}{2} + \epsilon}(y \tau,\sigma^{-1}\sqrt{y(1-y)[k^{2} + \sigma^{2}\tau^{2}]})\Bigl]\nonumber\\
&& + \frac{\epsilon}{2}F_{\frac{\epsilon}{2},1}^{(\tau)}(k,i=0,\sigma)\Bigr).
\end{eqnarray}
Next, use the property $\Gamma(x+1)=x \Gamma(x)$ in order to get a result 
purely in terms of $\epsilon$ when higher order terms are neglected. We then 
have 
\begin{eqnarray}\label{B12}
&& I_{3}^{(\tau)} (k,i=0;\sigma) =  \frac{1}{\epsilon}
\Bigl(-\frac{1}{4}(1 + \epsilon) 
\int_{0}^{1} dy (1-y)^{\frac{\epsilon}{2}-1}
\Bigr[\Bigr(y(1-y)[k^{2}\nonumber\\ 
&& + \sigma^{2}\tau^{2}]\Bigl)^{1-\epsilon} - \epsilon \sigma^{2-2\epsilon}f_{-\frac{1}{2} + \epsilon}(y \tau,\sigma^{-1}\sqrt{y(1-y)[k^{2} + \sigma^{2}\tau^{2}]})\Bigl]\nonumber\\
&& + \;\frac{\epsilon}{2}F_{\frac{\epsilon}{2},1}^{(\tau)}(k,i=0,\sigma)\Bigr).
\end{eqnarray}
We introduce two new parametric 
functions which will be useful to our future manipulations, whose expressions 
are given by
\begin{subequations}\label{B13}
\begin{eqnarray}
& \tilde{F}_{\alpha}^{(\tau)}(k,i=0;\sigma) = \sigma^{2-2\alpha}
\int_{0}^{1} dy (1-y)^{\frac{\alpha}{2}-1}
f_{-\frac{1}{2}+\alpha}(y \tau, \sigma^{-1} \sqrt{y(1-y)(k^{2} 
+ \sigma^{2} \tau^{2})}),\label{B13a}\\
& \bar{F}_{\alpha}^{(\tau)}(k,i=0;\sigma) = \sigma^{-2\alpha}
\int_{0}^{1} dy y (1-y)^{\frac{\alpha}{2}}
f_{\frac{1}{2}+\alpha}(y \tau, \sigma^{-1} \sqrt{y(1-y)(k^{2} 
+ \sigma^{2} \tau^{2})}).\label{B13b}
\end{eqnarray}
\end{subequations}
In terms of these functions, we can express the solution for the integral 
$I_{3}(k, i=0;\sigma)$ in a form appropriate to minimal subtraction, namely
\begin{eqnarray}\label{B14}
&& I_{3}^{(\tau)} (k,\sigma)= -\frac{1}{8\epsilon}
\bigl((k^{2} + \sigma^{2} \tau^{2})[1+\frac{\epsilon}{4} - 2\epsilon L_{3}(k^{2} + \sigma^{2} \tau^{2})]
- 2 \epsilon \tilde{F}_{\epsilon}^{(\tau)}(k,i=0;\sigma)\nonumber\\
&& - 4 \epsilon 
F_{\frac{\epsilon}{2},1}^{(\tau)}(k, i=0; \sigma)\bigr). 
\end{eqnarray}
If we prefer to employ the normalization condition scheme, we need the 
derivative of this integral with respect to $k^{2}$ computed at the symmetry 
point $k^{2}=1$. Henceforth we denote the argument of a function  
$(k,i=0,\sigma)$ at $k^{2}=1$ by $(\sigma)$. Using the recursion 
relation for Bessel functions
\begin{equation}\label{B15}
z^{-1}\frac{\partial\Bigl[z^{-\alpha}K_{\alpha}(\beta z)\Bigr]}{\partial z} = 
- \beta z^{-\alpha-1}K_{\alpha+1}(\beta z),
\end{equation}
and employing the explicit representation of $f_{\alpha}(a,b)$ in terms of the 
product involving the Bessel function and cosine, we learn that
\begin{equation}\label{B16}
I_{3SP}^{\prime (\tau)} (\sigma)= -\frac{1}{8\epsilon}
\bigl(1+\frac{5\epsilon}{4} - 2\epsilon ln(1+ \sigma^{2} \tau^{2})
+ 2 \epsilon \bar{F}_{\epsilon}^{(\tau)}(\sigma)- 4 \epsilon 
F_{\frac{\epsilon}{2}}^{\prime (\tau)}(\sigma)\bigr). 
\end{equation}
\par  Let us conclude our analysis of the two-point function by calculating 
the three-loop integral $I_{5}^{(\tau)}(k,i=0;\sigma)$. First, write it as 
\begin{eqnarray}\label{B17}
&& I^{(\tau)}_{5}(k,i=0;\sigma) = \sigma \sum_{j=-\infty}^{\infty} 
\int d^{d-1}q \frac{[I_{2}^{(\tau)}(q,j;\sigma)]^{2}}
{[(q-k)^{2} + \sigma^{2}(j+\tau)^{2}]}. 
\end{eqnarray} 
When we use the explicit expression for the four-point one-loop subdiagram 
along with the value of the integral over the parameter originally appearing 
there, expanding the $\Gamma$ functions in $\epsilon$ and neglecting higher 
order terms, we obtain
\begin{eqnarray}\label{B18}
& I_{5}^{(\tau)} (k,i=0;\sigma) =  \frac{1}{\epsilon^{2}}
\Bigl(\sigma (1+\epsilon) 
\sum_{j=-\infty}^{\infty} 
\int \frac{d^{d-1}q}{[(q-k)^{2} + \sigma^{2}(j+\tau)^{2}]
[q^{2} + \sigma^{2}j^{2}]^{\epsilon}} \nonumber\\
& + \epsilon \sigma
\int \frac{d^{d-1}q}{[(q-k)^{2} + \sigma^{2}(j+\tau)^{2}+1]} F^{(\tau)}_{\frac{\epsilon}{2}}(q,i=0;\sigma)\Bigr).
\end{eqnarray} 
Using completely similar steps in the calculation of 
$I_{3}^{(\tau)}(k,i=0;\sigma)$ with minor modifications, we can proceed 
henceforth quite analogously. Define the functions
\begin{subequations}\label{B19}
\begin{eqnarray}
& \check{F}_{\frac{3\epsilon}{2}}^{(\tau)}(k,i=0;\sigma) = 
\sigma^{2-3\epsilon}
\int_{0}^{1} dy (1-y)^{\epsilon-1}
f_{-\frac{1}{2}+\frac{3\epsilon}{2}}(y \tau, \sigma^{-1} \sqrt{y(1-y)(k^{2} 
+ \sigma^{2} \tau^{2})}),\label{B19a}\\
& \hat{F}_{\frac{3\epsilon}{2}}^{(\tau)}(k,i=0;\sigma) = \sigma^{-3\epsilon}
\int_{0}^{1} dy y (1-y)^{\epsilon} f_{\frac{1}{2}+\frac{3\epsilon}{2}}
(y \tau, \sigma^{-1} \sqrt{y(1-y)(k^{2} 
+ \sigma^{2} \tau^{2})}).\label{B19b}
\end{eqnarray}
\end{subequations}
The solution for $I_{5}$ useful in utilizing minimal subtraction scheme can 
be written as
\begin{eqnarray}\label{B20}
&& I_{5}^{(\tau)} (k,\sigma)= -\frac{1}{6\epsilon^{2}}
\bigl((k^{2} + \sigma^{2} \tau^{2})[1+\frac{\epsilon}{2} - 3\epsilon L_{3}(k^{2} + \sigma^{2} \tau^{2})]
- 3 \epsilon \check{F}_{\frac{3\epsilon}{2}}^{(\tau)}(k,i=0;\sigma)\nonumber\\
&&  - 6 \epsilon F_{\frac{\epsilon}{2},1}^{(\tau)}(k, i=0; \sigma)\bigr). 
\end{eqnarray}
Instead, if we wish to employ normalization conditions, we have to compute 
the derivative in relation to $k^{2}$ at the symmetry 
point $k^{2}=1$ and we get to  
\begin{equation}\label{B21}
I_{5SP}^{\prime (\tau)} (\sigma)= -\frac{1}{6\epsilon^{2}}
\Bigl(1+ 2\epsilon - 3\epsilon \Bigl[\frac{1}{2}ln(1+ \sigma^{2} \tau^{2})
- \hat{F}_{\frac{3\epsilon}{2}}^{(\tau)}(\sigma)+ 2  
F_{\frac{\epsilon}{2}}^{\prime (\tau)}(\sigma)\Bigr]\Bigr). 
\end{equation}
\par It is worthy to mention at this point that the finite size corrections 
for $I_{3SP}^{\prime (\tau)} (\sigma)$ and $I_{5SP}^{\prime (\tau)} (\sigma)$ 
become simpler when computed at $\epsilon=0$. In fact, defining the massless 
quantity $W_{0}^{(\tau)}(\sigma)= \frac{1}{2}ln\bigl(1+ \sigma^{2}\tau^{2}\bigr) 
- \hat{F}_{0}^{(\tau)}(\sigma) + 2 F_{0}^{\prime (\tau)}(\sigma)$, those 
integrals can be written simply as
\begin{subequations}
\begin{eqnarray}\label{B22}
&& I_{3SP}^{\prime (\tau)} (\sigma)= -\frac{1}{8\epsilon}
\bigl(1+\frac{5\epsilon}{4} - 2\epsilon W_{0}(\sigma)\bigr),\label{B22a}\\
&& I_{5SP}^{\prime (\tau)} (\sigma)= -\frac{1}{6\epsilon^{2}}
\bigl(1+ 2\epsilon - 3\epsilon W_{0}(\sigma)\Bigr). \label{B22b} 
\end{eqnarray}
\end{subequations}
\par The massless counterpart of the integral 
which contributes to the four-point function at two-loops can be written in 
terms of $I_{2}$ and reads:
\begin{equation}\label{B23}
I_{4}^{(\tau)}(k_{i},i_{i};\sigma) = \sigma \sum_{j=-\infty}^{\infty} 
\int d^{d-1}q  \frac{I_{2}^{(\tau)}(q+k_{3},j+i_{3};\sigma)}
{[(q)^{2} + \sigma^{2}(j+\tau)^{2}][(q-P)^{2} + \sigma^{2}(j-p+\tau)^{2}] }.
\end{equation} 
Once again we shall restrict ourselves to the simplest expressions for this 
integrals, namely those with 
zero mode ($p=0$) characterizing the external quasimomentum associated to the 
finite size direction, perpendicular to the plate surfaces. At the symmetry 
point, all primitively divergent vertex parts depend on only one external 
momenta scale. For this reason we are going to list the last integral at a 
certain external momentum $P$ and display the results in the most convenient 
form either using minimal subtraction or normalization conditions 
renormalization schemes. 
\par  We solve the internal bubble, i.e., we compute the integral over the 
momenta. Then, use a Feynman parameter $x$ to melt the integer powers of the 
propagators in a single denominator. We are left with an integral over the 
Feynman parameter $x$ as follows
\begin{eqnarray}\label{B24}
&& I_{4}^{(\tau)}(k_{i},0;\sigma) = [\frac{1}{\epsilon}(1+\frac{\epsilon}{2})
\sigma \int_{0}^{1} dx \sum_{j=-\infty}^{\infty} \int \frac{d^{d-1}q} 
{[q^{2} - 2xPq + x P^{2} + \sigma^{2}(j+\tau)^{2}]^{2}}\nonumber\\
&&\;\times 
\frac{1}{[(q+k_{3})^{2} + \sigma^{2}j^{2}]^{\frac{\epsilon}{2}}} 
+ \frac{1}{2}G_{\frac{\epsilon}{2}}^{(\tau)}(P,k_{3},0,\sigma) S_{d}],
\end{eqnarray}
where 
\begin{equation}\label{B25}
G_{\alpha}^{(\tau)}(P,k_{3},0,\sigma) = \frac{\sigma}{S_{d}} \sum_{j=-\infty}^{\infty} 
\int \frac{d^{d-1}q F_{\alpha}(q+k_{3},j,\sigma)}{[(q-P)^{2} + 
\sigma^{2}(j+\tau)^{2}][q^{2} + \sigma^{2}(j+\tau)^{2}]}.
\end{equation}
\par We shall be concerned with the behavior of $F_{\alpha}(q,j,\sigma)$ in the limit $q \rightarrow \infty$. In that limit, if this object behaves as an 
inverse power of $q$ all is well, since last integral will have no dimensional 
poles in $\epsilon$ using simple power counting arguments. Therefore, it 
can be neglected in our consideration of poles from $I_{4}(k,i=0,\sigma)$. 
Let us turn our attention to this issue. From its explicit form 
$F_{\alpha}(q,j,\sigma)< F_{\alpha}(q,j=0,\sigma)$ and we just have to 
discuss the latter in the appropriate limit.
\par From our discussion in the main text Sec. IV, we already know that
\begin{eqnarray}\label{B26}
& F^{(\tau)}_{\frac{\epsilon}{2}}(k,i=0;\sigma) = 4k^{-\frac{\epsilon}{2}} 
\sigma^{\frac{\epsilon}{2}} 
\overset{\infty}{\underset{n=1}{\sum}} cos(2\pi n\tau )
(\pi n)^{\frac{\epsilon}{2}} \int_{0}^{1} dx [x(1-x)]^{-\frac{\epsilon}{4}}
\nonumber \\
& \times K_{\frac{\epsilon}{2}}(2 \pi n k \sigma^{-1}[x(1-x)]^{\frac{1}{2}}).
\end{eqnarray} 
The integrand is symetric around $x=\frac{1}{2}$. Then we can multiply the 
integral by two and use the integration limits at $(0,\frac{1}{2})$. Next, 
perform the change of variables $x(1-x)= \Bigl(\frac{y}{2}\Bigr)^{2}$. Thus, 
we can write the integral in the form
\begin{equation}\label{B27}
\int_{0}^{1} dx [x(1-x)]^{-\frac{\epsilon}{4}} K_{\frac{\epsilon}{2}}(2 \pi n k \sigma^{-1}[x(1-x)]^{\frac{1}{2}})= 2^{\frac{\epsilon}{2}}\int_{0}^{1} dy 
y^{1-\frac{\epsilon}{2}} (1-y^{2})^{-\frac{1}{2}}K_{\frac{\epsilon}{2}}(\pi n k \sigma^{-1} y) .
\end{equation}   
After some manipulation with special functions using Ref.\cite{GR}, last 
integral can be put in the form
\begin{eqnarray}\label{B28}
&& \int_{0}^{1} dx [x(1-x)]^{-\frac{\epsilon}{4}} K_{\frac{\epsilon}{2}}(2 \pi n k \sigma^{-1}[x(1-x)]^{\frac{1}{2}})= 2^{\frac{\epsilon}{2}} 
\frac{\pi \epsilon}{2\pi n k}cosec(\frac{\pi \epsilon}{2})\Bigl[cosh(\pi n k \sigma^{-1})\nonumber\\
&& \int_{0}^{\pi n k \sigma^{-1}}
\frac{sinht}{t} dt - sinh(\pi n k \sigma^{-1})\Bigl(\gamma 
+ ln(\pi n k \sigma^{-1}) + \int_{0}^{\pi n k \sigma^{-1}}
\frac{cosht-1}{t} dt\Bigr)\Bigr].
\end{eqnarray}
The first integral in the right-hand side is defined as the integral 
hyperbolic sine and denoted by $shi(\pi n k \sigma^{-1})$. The complete term 
multiplying $sinh(\pi n k \sigma^{-1})$ is defined as  the integral 
hyperbolic cosine whose symbol is $chi(\pi n k \sigma^{-1})$. Using the 
identity $cosec(\frac{\pi \epsilon}{2})= \frac{2}{\pi \epsilon} + \frac{\pi \epsilon}{12} + O(\epsilon^{2})$, the above expression becomes regular in the limit $\epsilon \rightarrow 0$. Now, the finite size contribution is 
$O(\epsilon^{0})$ (regular) and we can set $\epsilon=0$ inside its expression 
in order to obtain
\begin{eqnarray}\label{B29}
&& F^{(\tau)}_{\frac{\epsilon}{2}}(k,i=0;\sigma)= \frac{4\sigma}{\pi k}
\overset{\infty}{\underset{n=1}{\sum}} \frac{cos(2\pi n\tau )}{n}
[cosh(\pi n k \sigma^{-1})shi(\pi n k \sigma^{-1})\nonumber\\
&&  -sinh(\pi n k \sigma^{-1})
chi(\pi n k \sigma^{-1})] + O(\epsilon).
\end{eqnarray}
Consider the terms inside the brackets and define $z=\pi n k \sigma^{-1}$. For 
example, from our definitions, we can write
\begin{equation}\label{B30}
coshz shiz - sinhz chiz= \frac{e^{z}}{2}(shiz - chiz) + \frac{e^{-z}}{2}(shiz + chiz).
\end{equation}
We can work out these equations further in order to reduce them in terms of the incomplete Gamma function defined by $\Gamma(0,z)= \int_{z}^{\infty} dt 
\frac{e^{-t}}{t}$, such that 
\begin{equation}\label{B31}
coshz shiz - sinhz chiz= \frac{e^{z}}{2}\Gamma(0,z) 
+ \frac{e^{-z}}{2}\Gamma(0,-z).
\end{equation}
Now the limit $k \rightarrow \infty$ is the same as $z \rightarrow \infty$.   
From Ref.\cite{GR}, the asymptotic value for real z is given by 
${\underset{z\rightarrow}{lim}}\Gamma(0,z)=\frac{e^{-z}}{|z|}$. Therefore, 
we find 
\begin{equation}\label{B32}
{\underset{z\rightarrow}{lim}}(coshz shiz - sinhz chiz)= \frac{1}{z}.
\end{equation} 
Thus, we conclude that
\begin{equation}\label{B33}
{\underset{k\rightarrow}{lim}}F^{(\tau)}_{\frac{\epsilon}{2}}(k,i=0;\sigma)= 
\frac{4\sigma^{2}}{\pi^{2} k^{2}}\overset{\infty}{\underset{n=1}{\sum}} \frac{cos(2\pi n\tau )}{n^{2}},
\end{equation}
showing that this behavior guarantees that the integral 
$G_{\frac{\epsilon}{2}}^{(\tau)}(k,i=0,\sigma)$ is regular and can be 
neglected in the determination of the coefficients of the pole terms contained 
in this diagram.
\par In the following we outline the method of computation of 
$I_{4}^{(\tau)}(k,i=0,\sigma)$. It does not bring essential new features in 
comparison with the massive case, being actually simpler its evaluation. 
Consider the first term in the last expression $I_{4}^{(\tau)}(k,i=0,\sigma)$. 
Using an additional Feynman parameter $y$ 
to fold the remaining terms into a single denominator in order to compute 
the integral over the momenta $q$, the resulting parametric integral over $y$ 
has a pole in $y=1$. In analogy to the massive case in Appendix A, we keep 
the prefactor (depending only on $y$) in that 
integral, and set $y=1$ in the overall term which depends on $(x,y,P,\sigma)$. 
We then find 
\begin{equation}\label{B34}
I_{4}^{(\tau)} (k_{i},0;\sigma)= \frac{1}{2 \epsilon^{2}}
\bigl(1-\frac{\epsilon}{2}- \epsilon i(P) + \epsilon F_{\epsilon}^{(\tau)}
(\frac{PL}{2\pi}, 0)\bigr),
\end{equation}
which at the symmetry point $P^{2}=\kappa^{2}=1$ takes a simpler form and 
simplifies further if the finite size correction is computed at $\epsilon=0$,
namely
\begin{equation}\label{B35}
I_{4SP}^{(\tau)} (\sigma)= \frac{1}{2 \epsilon^{2}}
\bigl(1+\frac{3\epsilon}{2} + \epsilon F_{0}^{(\tau)}
(\sigma)\bigr).
\end{equation} 

We are going to use these results in Sections IV and V in the massless 
computation of the critical indices. Notice that the results are such 
that $F_{\alpha}^{(\tau)}(\sigma)$ used in the present appendix 
is the appropriate object for the massless theory, given in the text by 
Eq.(\ref{49}).

\end{document}